\documentclass[aps,prl,superscriptaddress,twocolumn,citeautoscript, longbibliography]{revtex4-1}
\usepackage{graphicx}
\usepackage{amsfonts}
\usepackage{amsmath}
\usepackage{amssymb}
\usepackage{bm}
\usepackage{textcomp}
\usepackage{verbatim}
\usepackage{xspace}
\usepackage{soul}
\usepackage{bbold}
\usepackage{mathtools}
\usepackage{dcolumn}
\usepackage{hyperref}
\usepackage[usenames]{color}
\usepackage{subfig}
\usepackage{physics}
\usepackage{units}
\usepackage[export]{adjustbox}
\usepackage{tablefootnote}
\usepackage[dvipsnames]{xcolor}
\usepackage{ragged2e}
\DeclareCaptionJustification{justified}{\justifying}
\def\beq{\begin{equation}}
\def\eeq{\end{equation}}

\newcommand{\bmro}{Ba$_2$MgReO$_6$ \xspace}
\newcommand{\Gm}{\Gamma}

\newcommand{\out}[1]{{}}

\newcommand{\kvec}[1]{{\mathbf{#1}}}

\newcommand{\xs}{$x^2 - y^2$ \xspace}
\newcommand{\zs}{$z^2$ \xspace}
\newcommand{\xy}{$xy$ \xspace}

\begin{document}

%\title{Hidden order controlled by the interplay between electronic exchange and vibronic effects in Ba$_2$MgReO$_6$ from first principles}
%Electron-lattice interactions and phase transitions in Ba$_2$MgReO$_6$}
\title{Interplay of superexchange and vibronic effects in the hidden order of Ba$_2$MgReO$_6$ from first principles}

\author{Dario Fiore Mosca}
\address{CPHT, CNRS, \'Ecole polytechnique, Institut Polytechnique de Paris, 91120 Palaiseau, France}
\address{Coll\`ege de France, Université PSL, 11 place Marcelin Berthelot, 75005 Paris, France}
% \address{University of Vienna, Faculty of Physics and Center for Computational Materials Science, Vienna, Austria}

\author{Cesare Franchini}
\address{University of Vienna, Faculty of Physics and Center for Computational Materials Science, Vienna, Austria}
\address{Department of Physics and Astronomy "Augusto Righi", Alma Mater Studiorum - Universit\`a di Bologna, Bologna, 40127 Italy}

\author{Leonid V. Pourovskii}
\address{CPHT, CNRS, \'Ecole polytechnique, Institut Polytechnique de Paris, 91120 Palaiseau, France}
\address{Coll\`ege de France, Université PSL, 11 place Marcelin Berthelot, 75005 Paris, France}

\begin{abstract}
%The cryptic quadrupolar and magnetic low-temperature orders observed in the spin-orbit double perovskite Ba$_2$MgReO$_6$ are classified as "hidden" signifying that their origin poses a challenge for explanation through standard experimental and theoretical techniques. 
The origin of the  "hidden" quadrupolar and unconventional magnetic low-temperature orders observed in the spin-orbit double perovskite Ba$_2$MgReO$_6$  defies  %challenge for 
explanation 
through 
standard experimental and theoretical techniques.
Here we address this problem by deriving and solving an ab initio low-temperature effective Hamiltonian including inter-site electronic exchange and vibronic (electron-lattice) couplings between $J_{\mathrm{eff}}=3/2$ Jahn-Teller-active Rhenium states. 
Our findings disclose the nature of these  elusive states, attributing it to 
%the entanglement of lattice and electronic degrees 
intertwined  exchange and electron-lattice couplings,
%of freedom
 thus diverging from the conventional dichotomy of purely electronic or lattice driving mechanisms.
Our results indicate the resilience of the quadrupolar hidden order under pressure, yet its rapid suppression under uniaxial strain suggests that external or lattice-induced distortions play a pivotal role in determining the relative stability of competing phases in Ba$_2$MgReO$_6$ and similar $d^1$ double perovskites.

\end{abstract}

%\date{\today}

\maketitle

{\it Introduction.} In recent years, correlated insulators with strong spin-orbit (SO) effect have become a fascinating area of study, revealing a variety of unconventional low-temperature orders and exotic states of matter~\cite{Witczak-Krempa2014,Takayama2021, Santini2009, Kuramoto2009}. Particularly intriguing are the heavy transition-metal compounds, which have been proposed to display a multitude of exotic phases, including the enigmatic Kitaev spin liquid of SO $d^5$ Mott insulators~\cite{Takagi2019,Jackeli2009}, and the complex high-rank magnetic and charge multipolar orders found in $d^1$ and $d^2$ SO double perovskites and halides~\cite{Maharaj2020, Lu2017,Hirai2020_PRR,Ishikawa2019,Paramekanti2020,Pourovskii2021}.

The discovery and understanding of these unconventional orders present significant challenges due to their hidden nature, which often escapes detection by traditional experimental probes~\cite{RevModPhys.83.1301}.

Initially, theoretical studies focused mainly on electronic exchange and electrostatic interactions as the primary forces driving the unconventional orders observed in $5d^1$ and $5d^2$ SO DP~\cite{Chen2010,Chen2011,Svoboda2021}. However, since the spin-orbit-entangled ground states in a cubic crystal field were identified as Jahn-Teller (JT) active~\cite{Takayama2021}, the potential role of electron-lattice (EL) interactions could no longer be excluded~\cite{Iwahara2018,Iwahara2023,Fiore_Mosca2021,soh2023spectroscopic}.  
This shift of perspective has added layers of complexity, as the relative strength and roles between EL effects and purely electronic couplings remains poorly understood. Gaining a thorough understanding of the correlation between JT effects and electronic exchange could provide access to physical regimes that are currently inaccessible and enhance our comprehension of many related SO transition metal systems.

% The interplay of EL and purely electronic couplings in real SO Mott insulators introduces additional layers of  complexities, since their relative strength and role in the unconventional orders are not well understood.
%In the case of spin-orbit TM compounds, the contributions of various couplings  are hard to disentangle experimentally, since   magnetic and Jahn-Teller-active charge degrees of freedom (DoF) can be simultaneously active and entangled within the ground-state  multiplets of TM ions.

%a comprehensive first-principles framework is lacking.
%A comprehensive theoretical picture is still lacking with .

In this context, a prototypical example is the Mott insulating double perovskite \bmro (BMRO here after). Experimental high-resolution x-ray %scattering
diffraction and resonant elastic x-ray scattering (REXS) resolve two phase transitions occurring upon lowering the temperature~\cite{Marjerrison2016,Takayama2021,Hirai2020_PRR, soh2023spectroscopic}. The first transition  at $T_q~\approx$~33~K is concomitant with a lattice distortion reducing the symmetry of the $Fm\Bar{3}m$ parent cubic structure. It is followed by the emergence of a canted antiferromagnetic (cAFM) phase at $T_m \approx$18~K with net magnetization along [110] axis.  Under applied pressure, the   high-$T$ phase is suppressed and, simultaneously, the cAFM order transforms to a pure AFM one~\cite{Arima2022}. Among various $d^1$ DP Ba$_2M$ReO$_6$ compounds, two phase transitions are observed for $M$=Mg, Zn, and Cd~\cite{Hirai2020_PRR,Hirai2021_JPCM,Barbosa2022}, but only a single AFM phase is found for $M$=Ca~\cite{Yammamura2006,Ishikawa2021}.

%In the seminal experimental paper of 
Hirai and coworkers~\cite{Hirai2020_PRR} assigned the transition at $T_q$ 
%was assigned 
to the onset of a purely quadrupolar phase consisting of \xs anti-ferroic (AF\xs)  and \zs ferroic components (F\zs), as inferred from the observed reduction of the lattice symmetry to tetragonal $P42/mnm$. 
% I WOULD REMOVE THIS, BUT DO AS YOU WISH (though Ref.~\cite{Lovesey2021} argued that the ferroic component is not consistent with the proposed space group). 
A similar picture -- of a high-$T$ quadrupolar AF\xs phase followed by the onset of a low-$T$ cAFM order --  was previously shown to emerge from intersite exchange  (IE) models supplemented by a sizeable electrostatic quadrupolar interaction ~\cite{Chen2010, Svoboda2021}. Inspired by these theories, Refs.~\cite{Hirai2020_PRR,Hirai2021_JPCM,Arima2022} attempted to construct a generic phase diagram for $d^1$ DP linking the disappearance  of quadrupolar AF\xs and cAFM phases in BMRO under pressure and their absence in Ba$_2$CaReO$_6$ to the volume dependence of superexchange and electrostatic quadupolar  couplings.

%But while theoretical studies have anticipated AF\xs phase, the F\zs configuration had not been postulated~\cite{Chen2010, Svoboda2021}. 
%Additionally,  a recent study finds the F\zs to be incongruent with the specific symmetry of the $T_q$ unit cell's structure~\cite{Lovesey2021}. 

%The origin of the quadrupolar order is also debated and shrouded in complexity. Early theories primarily linked it to the superexchange mechanism complemented by sizeable quadrupolar interactions~\cite{Chen2010, Svoboda2021}. However, these approaches overlook quadrupolar interactions among $t_{2g}$ charge operators, thereby predisposing the system to order the $e_{g}$ ones. Furthermore, first principles calculations  for BMRO and similar compounds indicate that neglecting EL prevents the formation of the observed ground state magnetic configuration~\cite{Tehrani2021, Fiore_Mosca2021}.

%A 
%more 
%recent work by 

Iwahara and colleagues~\cite{Iwahara2023} suggested an alternative EL-dominated model for $d^1$ DP and halides that is based on strong JT coupling between $e_g$ quadrupoles combined with a system independent IE term. They successfully reproduced both the composite quadrupolar order and cAFM phases, although for a specific set of parameters. 
%Their approach is based on strong JT coupling between $e_g$ quadrupoles combined with a system independent IE term.

The different highlighted schemes 
%show that 
thus reduce the problem to 
%either 
the contradictory purely electronic~\cite{Chen2010, Svoboda2021, Kubo2023} or EL-dominated limits~\cite{Iwahara2023}.  
%exposes contradictions.
First principles calculations for BMRO and analogous compounds suggest that  the observed 
ground-state 
magnetic order 
%is unstable in the 
cannot be explained without 
%including 
invoking
JT distortions~\cite{Tehrani2021, Fiore_Mosca2021}. Soh et al.~\cite{soh2023spectroscopic} estimated the energy of cooperative JT distortions in DFT finding it to totally dominate over electronic exchange  and to force the AF\xs order.
%Furthermore, first principles attempts to combine electronic-mediated and EL interactions for BMRO and analogous compounds suggest on one side that neglecting EL prevents the formation of the observed ground state magnetic configuration~\cite{Tehrani2021, Fiore_Mosca2021}, and on the other, as shown by Soh. and coworkers~\cite{soh2023spectroscopic}, that cooperative JT distortions dominate over electronic exchange and force the AF\xs order. 
Importantly, in all these ab initio studies JT dynamics has not been included.

The available literature thus hints at a delicate
balance between EL and IE mechanisms in BMRO 
%A comprehensive description of BMRO
that necessitates 
%the integration of electronic and vibronic effects on an
their treatment on the equal footing, validated by fully ab initio computations of materials-specific interaction parameters.

% However, the entanglement of lattice and electronic degrees of freedom
% in BMRO demonstrated by Soh using REXS and supported by an IEI effective Hamiltonian combined with multipolar density functional theory (DFT) for the lattice interactions calls for a revision of the existing models by integrating electronic and virbonic effect in the same framework and validated by materials specific interaction parameters computed fully ab initio.

%their assumptions regarding the relative magnitude, structure and volume dependence of those couplings rely on indirect measurements or simplified estimates and suffer from a lack of validation primarily due to the absence of explicit first-principles estimations. 
%Both interaction mechanisms need to be simulteneously included with all relevant coupling strength estimated from first principles to 
%identify the driving force behind the actual orderings occurring in BMRO  and to reliably predict phase diagrams as a function of external parameters. 
 %actual orderings occurring 
 %in BMRO and to 
 %the driving forces behind it. 

In this Letter, we propose 
%density functional theory(DFT)+dynamical mean-field theory (DMFT)~\cite{Georges1996,Anisimov1997_1,Lichtenstein_LDApp} in a quasi-atomic Hubbard-I (HI) \cite{hubbard_1} approximation with a force-theorem approach to intersite exchange interactions (IEI) \cite{Pourovskii2016} and density-functional perturbation theory~\cite{PhysRevB.73.045112,PhysRevB.43.7231} into a unified ab initio framework 
an advanced ab initio 
%methods in a unified 
framework for deriving a low-energy many-body effective Hamiltonian  of SO Mott insulators that fully includes both IE and EL interactions. Deriving such a many-body effective Hamiltonian for BMRO and solving it by a consistent single-site mean-field theory 
%By direct calculation of all interactions in our effective Hamiltonian,
we 
%successfully 
reproduce both the AF\xs and cAFM orders.
%and prove the absence of F\zs in the quadrupolar region.
% We find that neither IE nor EL terms alone can account for the observed quadrupolar 
%and experimental cAFM 
We find that BMRO is in a weak JT coupling limit and that neither IE nor EL terms alone can account for the observed quadrupolar 
order. 
%in BMRO. 
A faithful representation of the physics of BMRO becomes possible only by incorporating 
both 
%EL 
interactions. 
%in contrast to the assumptions of Ref.~\cite{Iwahara2023}, 
% We also find that BMRO is in a weak JT coupling limit, with both EL and IE terms evolving vs. volume and their changes largely compensating each other. 
% The overall picture - of the high-$T$ AF\xs and low-$T$ cAFM orders - remains remarkably unchanged upon volume varied over quite a broad range.
We also find upon varying volume that EL and IE terms  largely compensate each other.
The overall picture - of the high-$T$ AF\xs and low-$T$ cAFM orders - remains remarkably unchanged over quite a broad range.
%, contrary to the assumptions of Refs.~\cite{Hirai2020_PRR,Hirai2021_JPCM,Arima2022}.  
In contrast, small strains are predicted to rapidly suppress the quadrupolar order and induce a collinear AFM phase, thus explaining the findings of  Arima and coworkers by non-hydrostatic pressure conditions~\cite{Arima2022}.

%In addition, inspired by the recent experimental study of Arima and coworkers~\cite{Arima2022}, we examine the effect of pressure on the stability of the quadrupolar and magnetic phases in BMRO.  Our analysis reveals that the application  of pressure does not induce transitions towards competing charge or magnetic phases.  We prove that the suppression of the AF\xs phase under pressures greater than 7 GPa, alongside the emergence of collinear AFM phase, can be attributed to non-hydrostatic factors such as strain, rather than hydrostatic pressure alone. 
%Overall, our findings shed light on the mechanism of formation of complex multipolar orders, paving the way for a more complete understanding of SO correlated materials. 

{\it Effective Hamiltonian.}  In the 5d$^1$ cubic DP, the ground state multiplet is determined by the combined action of crystal field and SO coupling. The former, splitting the 5d orbitals in $e_g$ and $t_{2g}$ states, produces an effective orbital momentum $l = 1$ which couples to the spin $S = 1/2$, promoting the $J_{\mathrm{eff}} = 3/2$ SO-coupled ground state wavefunctions.  
The full microscopic many-body effective Hamiltonian contains three components, an electronic part, the local JT activity and the intersite elastic couplings (EC):
\beq
H_{\mathrm{eff}} = H_{IEI} + H_{JT} + H_{EC}.
\label{eq:hamiltonian}
\eeq

The electronic-mediated IE interactions (IEI) between the multipolar moments with defined total angular momentum $J = 3/2$ are described by the effective Hamiltonian  
\beq\label{eq:H_IEI}
    H_{IEI} = \sum_{\langle ij \rangle} \sum_{K,K', Q, Q'} V_{KK'}^{QQ'} (ij) O_{K}^{Q} (i) O_{K'}^{Q'} (j)
\eeq
where the first summation ($ij$) runs over the
%nearest-neighbor 
Re-Re bonds, the second one over the multipolar momenta of the ranks $K$,$K'$=1, 2, 3 and projections $Q$,$Q'=-K...K$,  $O_{K}^{Q}(i)$ is the normalized multipolar operator \cite{Santini2009} acting on the site $i$,  $V_{KK'}^{QQ'}(ij)$ represents the corresponding IEI. By symmetry analysis of the $O_h$ subductions relative to the two-site (bond) cluster invariant space group $D_{2h}$, one can identify all non-zero matrix elements of  $V_{KK'}^{QQ'}(ij)$. We have verified that our \emph{ab initio} computed IEI matrices agree with the form expected from this symmetry analysis. See Supplemental Material (SM)~\cite{supplmat} for further details.

The $J_{\mathrm{eff}} = 3/2$ multiplet is JT active with the corresponding quadrupole moments  coupled to the irreducible distortions $\{T_{2g}, E_{g} \} $  of the ReO$_6$ octahedron via
\beq
    H_{JT} = \sum_{i, \Gm} \sum_{Q \in \Gm} \frac{p^{2}_{Q} (i)}{2M_{\Gm}} + \Phi_{\Gm} \frac{q^2_{\Gm Q} (i) }{2} - g_{\Gm} O^{Q}_2 (i) q_{\Gm Q} (i) 
\eeq
where $\Gm \in \{T_{2g}, E_{g} \} $, while $Q$ runs over the distortion modes belonging to $\Gm$. 
Here $\Phi_{\Gm}$ is the diagonal term of the irreducible  force-constant matrix, $q_{\Gm Q} (i)$ is the distortion associated with the irreducible mode, $p_Q(i)$ and $M_{\Gm}$ are the corresponding momentum and reduced mass, respectively,  $g_{\Gm}$ is the JT coupling constant that couples the distortion $q_{\Gm Q} (i)$ to the electronic quadrupole operator $O_2^Q(i)$. 

Lastly, JT distortions on neighboring octahedral centers interact with each other through elastic couplings
\beq\label{eq:H_IEC}
    H_{EC} = \sum_{\langle ij \rangle} \sum_{\Gm \Gm' } \sum_{Q\in\Gm}\sum_{Q'\in\Gm'}\Phi_{\Gm \Gm'}^{QQ'} (ij) q_{\Gm Q} (i) q_{\Gm' Q'} (j) , 
\eeq 
where the summation ($ij$) goes over 
%nearest-neighbors (nn) and next nearest-neighbors (nnn) 
 Re pairs, and  $\Phi_{\Gm \Gm'}^{QQ'} $ is the irreducible force constant matrix coupling  distortions of the oxygen cage at the sites $i$ and $j$.

\begin{figure}[!tb]
  	\begin{centering}
  	\includegraphics[width=0.96\columnwidth]{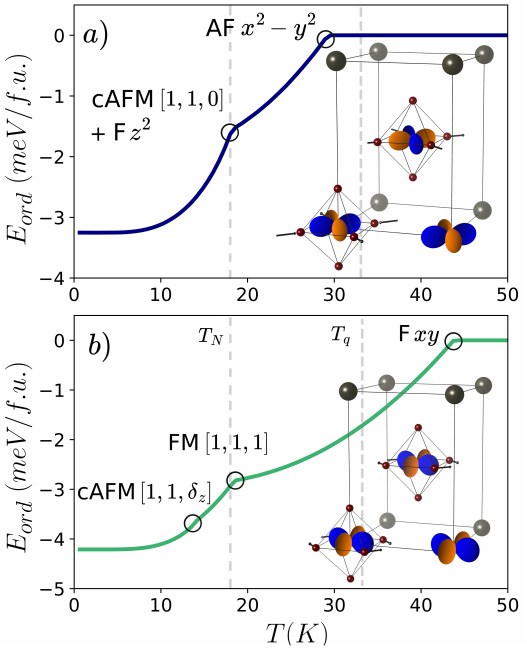} %pdf} 
  		\par\end{centering}
  	\caption{Mean-field ordering energy vs. temperature %solutions
   calculated from the Hamiltonian eq.~\ref{eq:hamiltonian}  with (a) all interactions included and (b) from $H_{IEI}$ only. The insets  depict the AF\xs (a) Fxy (b) ordered phases respectively, with associated JT distortions. The dashed lines mark the magnetic and quadrupolar measured temperatures~\cite{Hirai2020_PRR}.} 
  	\label{fig:E_MF} 
\end{figure}

{\it Methods.}   We determine all coupling constants in Eq. ~\ref{eq:hamiltonian} 
%we use a combination of first principles methods
from first principles.  The IEI are computed by the force-theorem approach of Ref.~\cite{Pourovskii2016}
%, which allows to calculate the IEI 
from the converged  paramagnetic electronic structure of cubic BMRO obtained
%from the solution of a fully converged 
by charge self-consistent density-functional theory (DFT)~\cite{Wien2k} + dynamical mean-field theory ~\cite{Georges1996,Anisimov1997_1,Lichtenstein_LDApp,Aichhorn2016}  in the quasi-atomic Hubbard-I (HI)~\cite{hubbard_1} approximation (see the SM ~\cite{supplmat}) . Our DFT+HI calculations correctly reproduce the Re$^{6+}$ ground state multiplet $J = 3/2$, the corresponding SO splitting between $J = 3/2$ and $J = 1/2$ states ($\approx 0.44$ eV) and the crystal field splitting ($\approx 4.5$ eV). The significant SO splitting supports the assumption of restricting the IEI to the ground state quartet, in contrast to other proposed approaches~\cite{qiu2021, Zhou2006, wang2017}. Its magnitude prevents strong inter-level mixing and is also substantially greater than the mean field ordering energy of approximately 4 meV (See Figure~\ref{fig:E_MF} a).  

The JT coupling constants $g_{\Gamma}$ are also obtained with DFT+HI by calculating a set of distorted BMRO structures for a chosen mode that is consistent with the order parameter and induces local octahedral distortions $Q \in \Gm$. We then  extract $g_{\Gm}$ by fitting the calculated $J = 3/2$ level splitting as a function of the mode amplitude (see the SM~\cite{supplmat}).
%for two distortion modes ($\Gm_{5+}, X_{2+}$) coupled to ($T_{2g}$-ferro, $E_g$-anti-ferro) quadrupolar orders respectively (See Supplemental~\cite{supplmat}).

Both on-site elastic %contribution to
constant in $H_{JT}$ and inter-site couplings in $H_{EC}$ have been obtained from the force constant matrix of a density functional perturbation theory calculation~\cite{PhysRevB.73.045112,PhysRevB.43.7231} projected onto the irreducible JT modes of the ligand ions via
\beq
      \Phi_{\Gm \Gm'}^{Q Q'} (ij) = \sum_{\alpha, \beta} \big(C_{\Gm Q}^{\alpha} \big)^\dagger   \Phi_{\alpha \beta} (ij)  \big( C_{\Gm' Q'}^{\beta} \big) \ , 
\eeq
where $\Phi_{\alpha \beta} (ij)$ is the force constant matrix in the Cartesian coordinate reference frame of components ($\alpha, \beta$) for atomic sites ($i$, $j$). The projected force constants are obtained through change of coordinate on the ligand ions $r_{\Gm Q} = \sum_{\alpha } C_{\Gm Q}^{\alpha} \ r_{\alpha}$. 

We note that our calculations predict only the nearest-neighbor IEI terms in Eq.~\ref{eq:H_IEI} to be substantial in BMRO. The longer-range IEI terms are deemed negligible, aligning with expectations for superexchange and consistent with our prior findings.~\cite{Fiore_Mosca2021,Pourovskii2021,Pourovskii2023}.
% that only nearest-neighbor IEI in Eq.~\ref{eq:H_IEI} are predicted by our calculations to be significant in BMRO, the longer-range IEI are negligible as expected for superexchange and in agreement with our previous results on IEI in SO DP 
The inclusion of next-nearest neighbors EC in Eq.~\ref{eq:H_IEC} is necessary for reproducing the phonon spectra at the commensurate points with good accuracy (see SM~\cite{supplmat}).

{\it JT regime.} Our calculations predict that $E_g$ modes are almost 3 times more strongly JT-coupled than the $T_{2g}$ ones, 
%the corresponding coupling values 
$g_{E_{g}} = 1.53$ eV/\AA~and $g_{T_{2g}} = 0.58$ eV/\AA. The resulting JT stabilization energy (i.~e. the depth of the JT potential well) $E_{JT} = g_{E_g}/ 8\Phi_{E_g} \approx 14$ meV,  well below the oscillatory frequency of the $E_g$ modes $\hbar \omega_{E_g}~\approx~75$ meV,
places BMRO in the weak coupling regime, in contrast to previous assumptions suggesting a strong coupling regime \cite{Iwahara2023}. %($\approx 50$ meV~~\cite{Iwahara2023}).
 % Our data places BMRO 
 % %is found to be 
 % in the weak JT coupling regime~\cite{bersuker_vibronic_1989}, 
 % in contrast with the strong regime assumed  in Ref.~\cite{Iwahara2023}.
 % Our computed JT energy $E_{JT} = g_{E_g}/ 8\Phi_{E_g} \approx 14$ meV, significantly smaller than previous estimation ($\approx 50$ meV~~\cite{Iwahara2023}) and  well below the oscillatory frequency of the $E_g$ modes $\hbar \omega_{E_gg}~\approx~75$ meV
 % In contrast, Ref.~\cite{Iwahara2023}  estimated the JT energy to be $\approx 50$ meV, suggesting a strong JT regime.  
 Our numerical tests on the single-site JT problem indeed show that the vibronic basis cut to one phonon -- corresponding to a weak-coupling treatment of the JT problem \cite{bersuker_vibronic_1989} -- already provides reasonably high accuracy for the ground-state energy (see SM~\cite{supplmat}).

\begin{figure}[!tb]
  	\begin{centering}
  	\includegraphics[width=0.99\columnwidth]{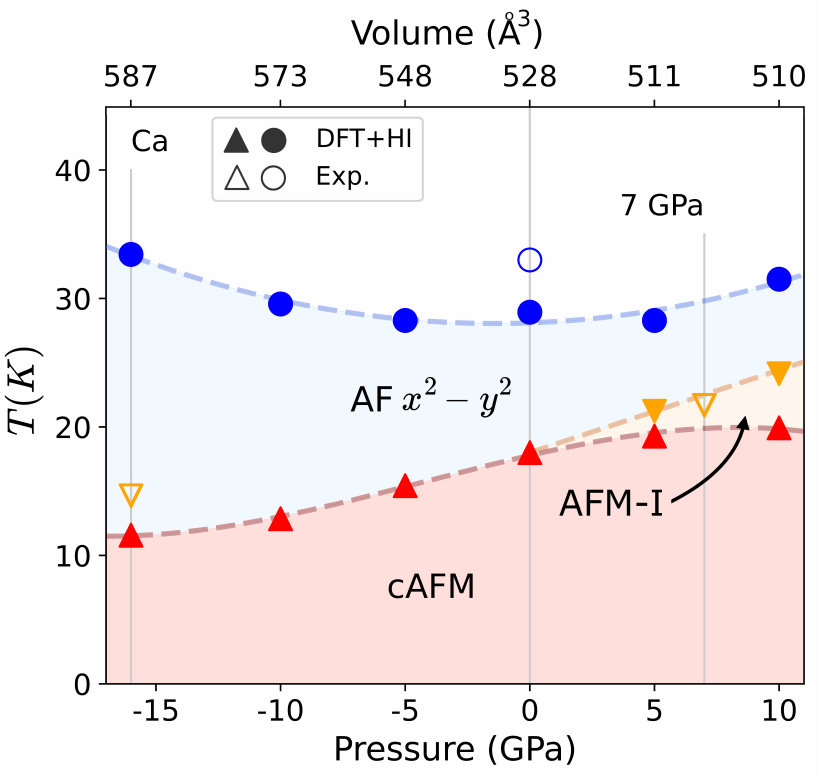} %pdf} 
  		\par\end{centering}
  	\caption{Mean-field phase diagram of BMRO at different pressures. The full circles and triangles represent the computed points, while the empty are experimental results. The 7 GPa exp. is from Ref.~\cite{Arima2022}, the -16 Gpa point corresponds to the Ca compound of Ref.~\cite{Yammamura2006}.  Upright and inverted triangles mark the cAFM [110], AFM-I  transition temperatures, while circles mark the AF\xs phase.} 
  	\label{fig:main2} 
\end{figure}

{\it Ordered phases.} As a next step we solve the Eq.~\ref{eq:hamiltonian} within a single-site mean field (MF)~\cite{Rotter2004} by decoupling both the IEI (Eq.~\ref{eq:H_IEI}) and EC (Eq.~\ref{eq:H_IEC}) terms and solving the single-site JT problem in mean-field in the weak-coupling limit. We find that BMRO undergoes two second-order phase transitions at temperatures $T_q \approx $  45 K and $T_m \approx$ 28 K. The calculated values are systematically overestimated by $\sim 40 \%$ as a consequence of the MF + HI approximation~\cite{Horvat2017,Pourovskii2019,Pourovskii2021}. %To reconcile this discrepancy, 
We have thus reduced the transition temperatures in our phase diagrams (Figures~\ref{fig:E_MF} and ~\ref{fig:main2}) by this constant factor. Our findings align well with experimental observations, showing a charge quadrupolar  AF\xs order with wave vector  
%$\kvec{k}=[0, 0, 1/2]$ 
$\kvec{k}=[0, 0, 1]$ 
below $T_q$, and cAFM configuration with the net moment along [110] below $T_m$. {Both the canting angle ($\phi = 30^\circ$) and the magnitude of the net moment ($\mu \approx 0.25 \mu_B$) compare very well with the experimental values of $\phi = 40^\circ$ and  $\mu \approx 0.3 \mu_B$~\cite{Hirai2020_PRR}.}
Our data does not reveal any ferroic F\zs in the quadrupolar region though it manifests below $T_m$.  
Its occurrence above $T_m$ in experiment might be due to high-order elastic couplings that are not accounted for in our model. It is worth noting that the measured magnitude of the \zs distortion at low temperatures is twice as large as the corresponding value at 25~K and does not appear to behave as a primary order parameter~\cite{Hirai2020_PRR}. Recent studies found that a strong intersite Coulomb
repulsion V might have an effect on the stability of the F\zs order~\cite{Zhang2024}.

% We note that in our calculations the ferroic F\zs is absent in the quadrupolar region, but emerges below $T_m$. This can be attribute to high-order elastic couplings not included in our model.
% This is consistent with the refinement of the  $I4/mmm$ structure at 6 K (see Supplementary of  Ref.~\cite{Hirai2020_PRR})\LP{Actually, this is not consistent, we need to rewrite that}. 
Simultaneously with F\zs, the $t_{2g}$ $xy$ charge quadrupole also orders ferroically below $T_m$. The corresponding amplitude of the F$xy$ distortion mode is more than an order of magnitude smaller than that of the AF\xs mode. In contrast, the ratio between \xs and \zs distortion modes is $\sim 4.6$, as compared to experimental ratio of $\sim 4.5$~\cite{Hirai2020_PRR}, though the predicted \xs distortion magnitude (0.021~\AA) is underestimated. 

In order to investigate competing orders and comprehend the influence of EL interactions on the stability of the observed AF\xs and cAFM phases, we conducted calculations with the JT and EC terms set to zero. %The outcomes presented in Fig.~\ref{fig:E_MF}
We obtained (Fig.~\ref{fig:E_MF}b) 
% To explore competing orders and understand the role of EL interactions on the stability of the observed AF\xs and cAFM phase, we repeated the calculation by setting the JT and IEC terms to zero.  The results displayed in Fig.~\ref{fig:E_MF}(b) demonstrate that vibronic interactions are crucial in determining the charge and magnetic order, particularly in establishing the observed AF\xs, which is not captured by considering only electronic interactions. Excluding EL interactions we obtain three phase transitions: 
(i) ferro-quadrupolar $xy$ (or equivalently $xz/yz$) phase (F$xy$) at temperature $T_q \approx 68$ K, followed by (ii) collinear ferromagnetic with [111] easy axis at $T_{FM} \approx 28$ K and (iii) cAFM with easy axis along $[1, 1, \delta_z]$ at $T_{m'} \approx 22$ K, with $\delta_z = 0.67$ as T$\to$0. 
% The role of EL interactions is thus crucial in determining the charge and magnetic order, particularly in establishing the observed AF\xs. 
%
This %observation
demonstrates that vibronic interactions play a crucial role in determining both charge and magnetic order, particularly in establishing the observed AF\xs state. 
The quadrupolar IEI alone, in the absence of local and non-local lattice interactions, favor ferro-quadrupolar F\xy, as follows from the IEI structure;
%can be elucidated by examining their structure: 
in fact the $t_{2g}$ IEI, which were overlooked in the electronic models of Refs.~\cite{Chen2010, Svoboda2021, Kubo2023}, exhibit a ferroic nature and are nearly equal in magnitude to the antiferroic $e_g$. Importantly, the ferroic order is not frustrated by the Re sublattice geometry.
%This observation is further supported by examining the relative energy scales involved.
% Through computing the single-site mean field energies across three different scenarios — one including only IEI, another with the JT term added, and the last incorporating IEC — 

To verify the essential role played by EC in stabilizing AF\xs phase we compared three different scenarios: 
(i) Considering only IEI, the total energy difference $\Delta E$ between %ferroelectric 
F\xy and AF\xs is -1.46 meV. (ii) Introducing the local JT coupling term reinforces the stability of F\xy, resulting in $\Delta E = -2.11$ meV. (iii) However, upon adding non-local EC, the energy balance is reversed, favoring the AF order with $\Delta E = 0.71$ meV.
Moreover, we also carried out calculations with only EL (the JT and EC terms) included. This results in a pure F$z^2$ order (that is  actually experimentally found in related d$^1$ halides \cite{Ishikawa2019}) with $T_q\approx$~20~K. Hence, neither the EL term alone can account for the correct QP order in BMRO.
% Through computing the single-site mean field energies across three different scenarios — one including only IEI, another with the JT term added, and the last incorporating IEC — we explicitly verified
% the crucial role of EIC in stabilizing AF\xs phase. When considering only IEI, the total energy difference $\Delta E$  between F\xy and AF\xs is -1.46 meV. Introducing the JT coupling term reinforces the stability of F\xy, with $\Delta E = -2.11$ meV.  When adding IEC, however, scenario is reversed with AF\xs order being favoured and $\Delta E = 0.71$ meV. 
  
\begin{figure}[!tb]
  	\begin{centering}
  	\includegraphics[width=1\columnwidth]{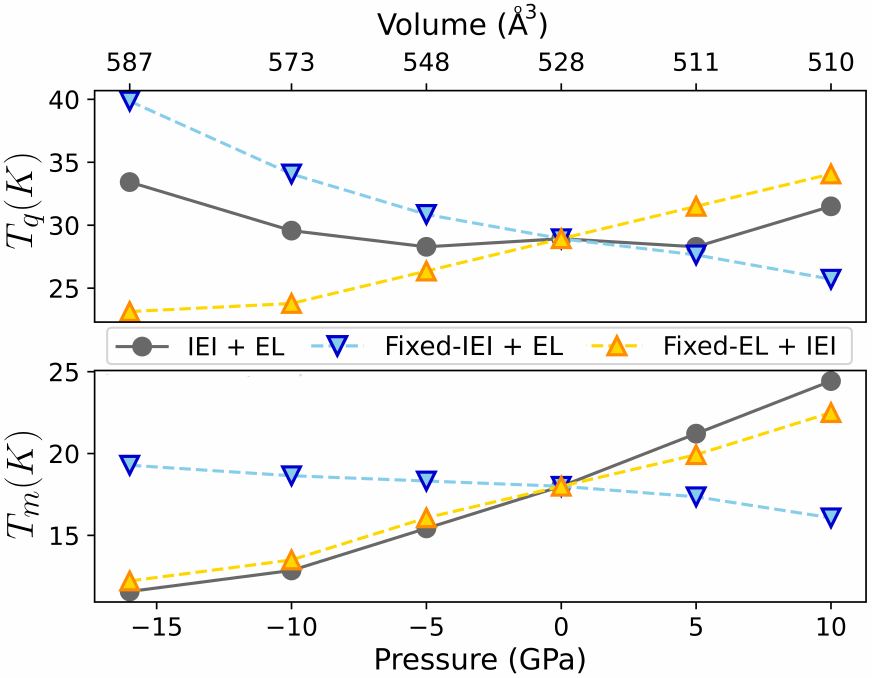} %pdf} 
  		\par\end{centering}
  	\caption{Mean-field transition temperatures dependence of $T_q$ (top) and $T_m$ (bottom), as a function of pressure for the full Hamiltonian of eq.~\ref{eq:hamiltonian} (filled circles). Together are plotted the solutions obtained with IEI fixed to 0 GPa, while varying EL interactions (inverted triangles) or with fixed EL, while varying IEI interactions (upright triangles).} 
  	\label{fig:gedanken} 
\end{figure}

{\it Effect of hydrostatic pressure.} 
%Pressure, both chemical and hydrostatic, can have a great impact on the stability of magnetic phases~\cite{}. 
%It was recently put forward by 
Arima and coworkers recently observed that pressures higher than $\approx$ 7 GPa suppress the AF\xs phase in BMRO and induce a cAFM to collinear AFM transition~\cite{Arima2022}. %At the same time, phase diagrams showing isostructural and isoelectronic compounds have proposed to connect the different magnetic phases based on volumetric effects~\cite{JPSJ.91.013702, Hirai_2021}.  
To understand the role of pressure on BMRO, we 
% have fitted a Birch-Murnaghan equation of state on DFT volumetric calculations~\cite{PhysRevB.54.11169,PhysRevB.47.558, DAR2018137}, see SM~\cite{supplmat} for details, and 
computed the IEI, JT and EC Hamiltonians as a function of pressure $P$ between -16 GPa (corresponding to the volume of Ba$_2$CaReO$_6$) and 10 GPa. 
Our results (Figure~\ref{fig:main2}) highlight the persistence of both AF\xs and cAFM phase at all pressures. For positive values, the AF\xs approaches the magnetic ordering, without any crossing and a small AFM region appears above the cAFM one. This AFM phase has a $2\kvec{k}$ structure with magnetic moments lying in the $xz$ plane (AFM-I). 
For negative pressures the $T_q$ increases, while the $T_m$ decreases. To decipher the peculiar behavior of quadrupolar and magnetic transition temperatures,  we have performed a \emph{gedankenexperiment} consisting in either keeping  IEI fixed at the 0 GPa values while tuning the EL interactions or vice versa. Our results (Fig.~\ref{fig:gedanken}) show  that the temperature behavior of $T_m$ is accurately captured using only IEI. However, regarding the quadrupolar phase, we note that while IEI alone would lead to a rise in the transition temperature as $P$ increases, the EL interactions counter act this effect by lowering it. In the end, it is the interplay between these two tendencies that shapes the observed temperature curve.

Therefore, our calculations 
reveal, in contrast to previous assumptions~\cite{Hirai2020_PRR,Hirai2021_JPCM,Arima2022}, that volume change does not qualitatively affect    
the phase stability in  d$^1$ Re DP.
Neither for the large-volume limit ($P$=-16 GPa, corresponding to Ba$_2$CaReO$_6$) nor under the highest pressure in Arima  measurements (10 GPa) we do find the experimentally observed collapse of the quadupolar order and onset of the collinear AFM. This apparent contradiction can be understood by considering non-isotropic pressure as elaborated in the following.
%This is clear by looking at the large-volume limit (corresponding to -16 GPa), which is the one of the isolectronic Ba$_2$CaReO$_6$, whose magnetic ground state configuration is collinear AFM. 
%We suggest that other effects can be important in this case, such as the lower structural symmetry of the system ($I4/m$) or external strain. Furthermore, we demonstrate the impact of the latter on the positive pressure side of the phase diagram. 

\begin{figure}[!b]
  	\begin{centering}
  	\includegraphics[width=0.99\columnwidth]{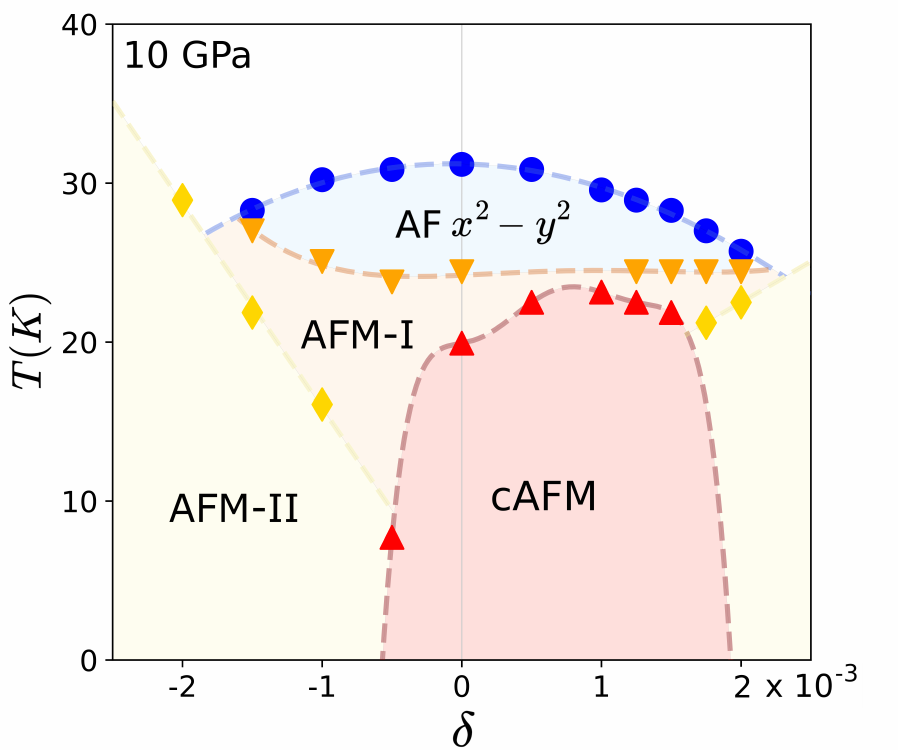} 
  		\par\end{centering}
  	\caption{Mean-field phase diagram showing the evolution of BMRO at 10 GPa under unaxial strain. Four phases can be distinguished: (i) AF\xs as circles, (ii) cAFM as upright triangles (iii) AFM-I as inverted triangles and (iv) AFM-II as diamonds. The curves are obtained from polynomial fits of the data points.} 
  	\label{fig:lepen} 
\end{figure}

%{\it Effect of non-hydrostatic pressure} 
{\it Effect of strain.} 
%We conclude by showing the effect strain on the 10 GPa solution. 
Glycerol pressure-transmitting medium used in Arima's experiment~\cite{Arima2022} does not ensure perfect hydrostaticity above 5 GPa, potentially leading to uniaxial stresses. To explore the impact of uniaxial strains in compressed BMRO, we introduced an on-site term to our effective Hamiltonian (Eq.~\ref{eq:hamiltonian}). This term is given by
%represented as 
$H_{\mathrm{strain}} = K \delta O_2^0 (i)$, with $\delta = c/a - 1$, where $K=$-3.0~eV was estimated from DFT+HI calculations for tetragonally distorted BMRO akin to the JT coupling.
% As pointed out by the authors of Ref.~\cite{Arima2022}, the glycerol pressure-transmitting media used in their experiment does not guarantee perfect hydrostaticity above 5 GPa, and uniaxial stresses can be expected.  
% We have investigated the effect of uniaxial strains in compressed BMRO by adding an on site term to the MBEH (\ref{eq:hamiltonian}) of the form $H_{\mathrm{strain}} = K \delta O_2^0 (i)$ with $\delta = c/a - 1$ and $K=$-3.0~eV estimated from DFT+HI calculations for tetragonaly distorted BMRO in the same way as the JT coupling. 

For $P$=10~GPa and the range $\delta \in [-2.5 \cdot 10^{-3}, 2.5 \cdot 10^{-3}]$, which corresponds to a tetragonal distortion of the unit cell of maximum $\pm$ 0.02~\AA, we obtain the phase diagram shown in Figure~\ref{fig:lepen}. It features four regions: AF\xs, cAFM [110], AFM-I and AFM-II, the last being a collinear AFM structure with wave vector $\kvec{k} = [0, 0, 1]$. 
While the cAFM and AFM-I MF solutions form asymmetric regions, 
the AF\xs phase is suppressed on the both sides for $|\delta|>2\cdot 10^{-3}$ in favour of the collinear AFM-I due to easy-plane(axis) single-ion anisotropy induced by tetragonal compression (elongation) by $H_{\mathrm{strain}}$. 
Our calculations thus predict that even small deviations from hydrostaticity will lead to the suppression of the quadrupolar order and appearance of a collinear AFM reported in Ref.~\cite{Arima2022}.

{\it Conclusions.} 
{We have derived an ab initio many-body effective Hamiltonian for BMRO, incorporating both electronic and vibronic interactions. Our evaluation of the ordered phases for BMRO under ambient conditions reveals that the experimentally
observed AF\xs quadrupolar order and non-collinear  canted AFM phases emerge from a subtle interplay of electronic exchange and electron-lattice interactions. In a first attempt to build a bridge towards a unified understanding of the entire $d^1$ double perovskite series, we find that the volume effect on phase stability is qualitatively insignificant. In contrast, even minute strains rapidly suppress the quadrupolar order and promote the stabilization of a collinear AFM state. 
Our results suggest that small purely lattice distortions, observed, for instance, in both Ba$_2$CaReO$_6$~\cite{Yammamura2006} and Ba$_2$CdReO$_6$~\cite{Hirai2021_JPCM}, rather than volume, can be identified as key parameter in controlling their low-temperature order. 
The proposed ab initio approach demonstrates broad applicability, suggesting its potential use in other Mott insulators with spin-orbit entangled and Jahn-Teller-active ground state multiplets, both in bulk and in the 2D limit. This method holds promise for  uncovering novel quantum states of matter and provide crucial theoretical insights into the underlying driving mechanisms.}

%To conclude, our first-principles calculations, encompassing both intersite exchange and electron-lattice interactions, successfully mimic the experimental AF\xs and cAFM phases observed in \bmro. This result is a consequence of the complex interplay of these interactions. Our findings underscore that Jahn-Teller and  intersite elastic couplings are instrumental in stabilizing the AF\xs phase. Neglecting electron-lattice terms would otherwise lead the system towards a competitive F\xy order. Furthermore, our pressure simulation results reveal the inherent robustness of the AF\xs and cAFM phases against external pressures, with only strain effects showing the potential to alter these states significantly.

\hspace{0.1cm}

\begin{acknowledgements}
We acknowledge useful discussions with N. Iwahara, M. Merkel and J.-R. Son.
Support by the the Austrian Science Fund (FWF) grant J4698 is gratefully  acknowledged.  
C.F. acknowledges financial support by  the European Union – Next Generation EU -
“PNRR - M4C2, investimento 1.1 - Fondo PRIN 2022” - “Superlattices of relativistic oxides” (ID 2022L28H97,
CUP D53D23002260006).
D.~F.~M. thanks C. Verdi and L. Ranalli for the useful discussions and acknowledges the computational facilities of the Vienna Scientific Cluster (VSC). L.~V.~P. is thankful to the CPHT computer team for support.
\end{acknowledgements}

\bibliography{bibliography}

\end{document}

\title{Supplementary material for 
'Interplay of superexchange and vibronic effects in the hidden order of Ba$_2$MgReO$_6$  from first principles'}

\author{Dario Fiore Mosca}
\address{CPHT, CNRS, \'Ecole polytechnique, Institut Polytechnique de Paris, 91120 Palaiseau, France}
\address{Coll\`ege de France, Université PSL, 11 place Marcelin Berthelot, 75005 Paris, France}

\author{Cesare Franchini}
\address{University of Vienna, Faculty of Physics and Center for Computational Materials Science, Vienna, Austria}
\address{Department of Physics and Astronomy "Augusto Righi", Alma Mater Studiorum - Universit\`a di Bologna, Bologna, 40127 Italy}

\author{Leonid V. Pourovskii}
\address{CPHT, CNRS, \'Ecole polytechnique, Institut Polytechnique de Paris, 91120 Palaiseau, France}
\address{Coll\`ege de France, Université PSL, 11 place Marcelin Berthelot, 75005 Paris, France}

\date{\today}	
\maketitle

\section{First principles methods}

In the following, we first describe our electronic structure calculations for paramagnetic BMRO. We then proceed with the description of how  the intersite exchange interaction (IEI) and Jahn-Teller (JT) terms of the full effective Hamiltonian (5) are calculated on the basis of this electronic structure. Lastly, we describe the intersite elastic coupling (EC) calculations by  the Density Functional Perturbation Theory (DFPT) approach.

\subsection{Correlated electronic structure calculations}\label{ssec:dft+hi}

 The electronic structure of Ba$_2$MgReO$_6$ in the paramagnetic phase is computed with the DFT+dynamical mean-field theory(DMFT) method. The quantum impurity problem for the $t_{2g}$ shell of the Re ion is solved within the quasi-atomic Hubbard-I (HI) approximation~\cite{hubbard_1}; the method is abbreviated below as DFT+HI.  We employ a full charge self-consistent DFT+DMFT implementation\cite{Aichhorn2009,Aichhorn2011,Aichhorn2016} based on the full-potential LAPW code Wien2k\cite{Wien2k} that includes the spin-orbit coupling effect with the standard variational treatment. 

The Wannier orbitals representing Re $t_{2g}$ orbitals are constructed from the manifold of $t_{2g}$ Kohn-Sham (KS) bands enclosed by the energy window [-1.36:2.04]~eV relative to the KS Fermi level. 
The Kanamori rotationally-invariant on-site Coulomb repulsion between Re $t_{2g}$ is parameterized by $U_{\mathrm{K}}=3.77$~eV and $J_{\mathrm{H}}^{\mathrm{K}}=0.39$~eV, which correspond to the full $d$-shell parameters $F^0=U=$3.2~eV and $J_{\mathrm{H}}$=0.5~eV. Those values of $U$ and $J_{\mathrm{H}}$ are in agreement with our previous works on $d^1$ and $d^2$ DP~\cite{Fiore_Mosca2021,Pourovskii2021}.

The calculations of Figure 1 in the main are carried out for the experimental cubic lattice structures of Ba$_2$MgReO$_6$ with lattice parameter $a = $ 8.0802 \AA~\cite{}. The local density approximation of the DFT exchange-correlation potential is employed, together with 400 $\vk$-point mesh in the full Brillouin zone, and the Wien2k basis cutoff $R_{\mathrm{mt}}K_{\mathrm{max}}=$7. The double-counting correction is evaluated using the fully-localized limit with the nominal 5$d$ shell occupancy of 1.

\subsection{Calculation H$_{IEI}$}

We evaluate the IEI from the converged electronic paramagnetic electronic structure of BMRO using the force-theorem in Hubbard-I (FT-HI) approach\cite{Pourovskii2016}. This method considers small symmetry-breaking fluctuations in the density matrix of the ground state (GS) $J_{\mathrm{eff}}=3/2$ multiplet that occur simultaneously on two neighboring magnetic (Re) sites $i$ and $j$. The IEI $V_{KK'}^{QQ'} (ij)$ is then evaluated by considering the response of DFT+DMFT grand potential to these two-site fluctuations. The method is similar in spirit to the force-theorem methods for the symmetry-broken magnetic state  of Refs.~\onlinecite{Liechtenstein1987,Katsnelson2000b}, but is formulated, instead, for the symmetry-unbroken paramagnetic state. The application of FT-HI in the case of BMRO is fully analogous to its previous applications to $J_{\mathrm{eff}}=3/2$ DP~\cite{Fiore_Mosca2021,Pourovskii2023}. The reader can thus find more detailed description in Appendix of Ref.~\onlinecite{Pourovskii2023} and  Suppelementary of Ref.~\onlinecite{Fiore_Mosca2021}; the full derivation of the method is given in Ref.~\onlinecite{Pourovskii2016}.

\subsection{Calculation $H_{JT}$}

\begin{figure}[!h]
    \centering
    \includegraphics[scale=0.3]{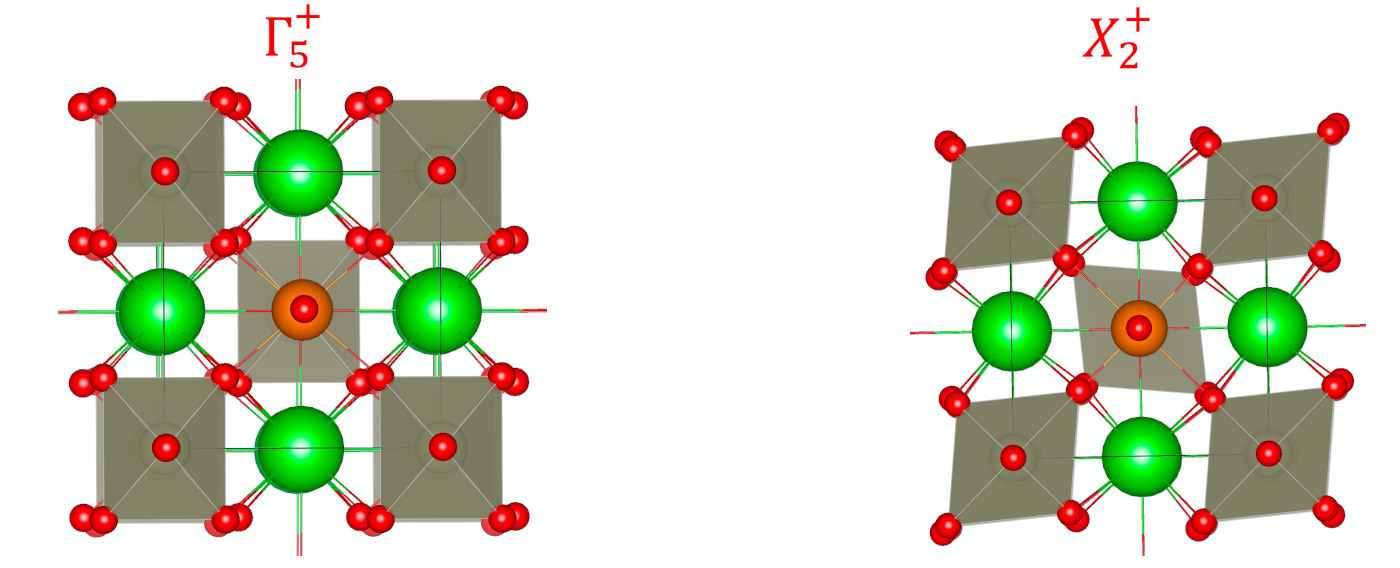}
    \caption{Distorted unit cells of BMRO implementing the $\Gm_5^+$ and $X_2^+$ distortions. The displayed distortion amplitude  is strongly exaggerated for better visibility as  compared to the distortions amplitude actually used in calculations. The plots are generated by VESTA \cite{VESTA}.}
    \label{fig:dist_struct}
\end{figure}

Since the distortions of ReO$_6$ octahedra couple to quadrupoles and we are interested  in the competition between the F\xy  and AF\xs  quadrupolar orders, we consider the corresponding oxygen sublattice distortions belonging to the Brillouin zone $\Gm$ and $X$-points irreducible representations that induce $T_{2g}$ and $E_{g}$ local distortions of the ReO$_6$ octahedra. The corresponding $\Gm_5^+$ and $X_2^+$ structures (see Suppl. Fig.~\ref{fig:dist_struct}) were generated with the help of ISODISTORT tool  \cite{campbell_isodisplace_2006,stokes_isodistort_2022} for a set of small distortions with the amplitude in the range from about 0.01~\AA\ to 0.03~\AA. 

Subsequently we carry out DFT+HI calculations using the same setup as described in Sec.~\ref{ssec:dft+hi}. In these self-consistent calculations, we follow the approach of Ref.~\onlinecite{Delange2017} by averaging the Boltzmann weights within the \Jeff quadruplet in order to suppress the unphysical self-interaction contribution\cite{brooks1997density} to the quadruplet's splitting. 

\begin{figure}[!h]
    \centering
    \includegraphics[scale=0.5]{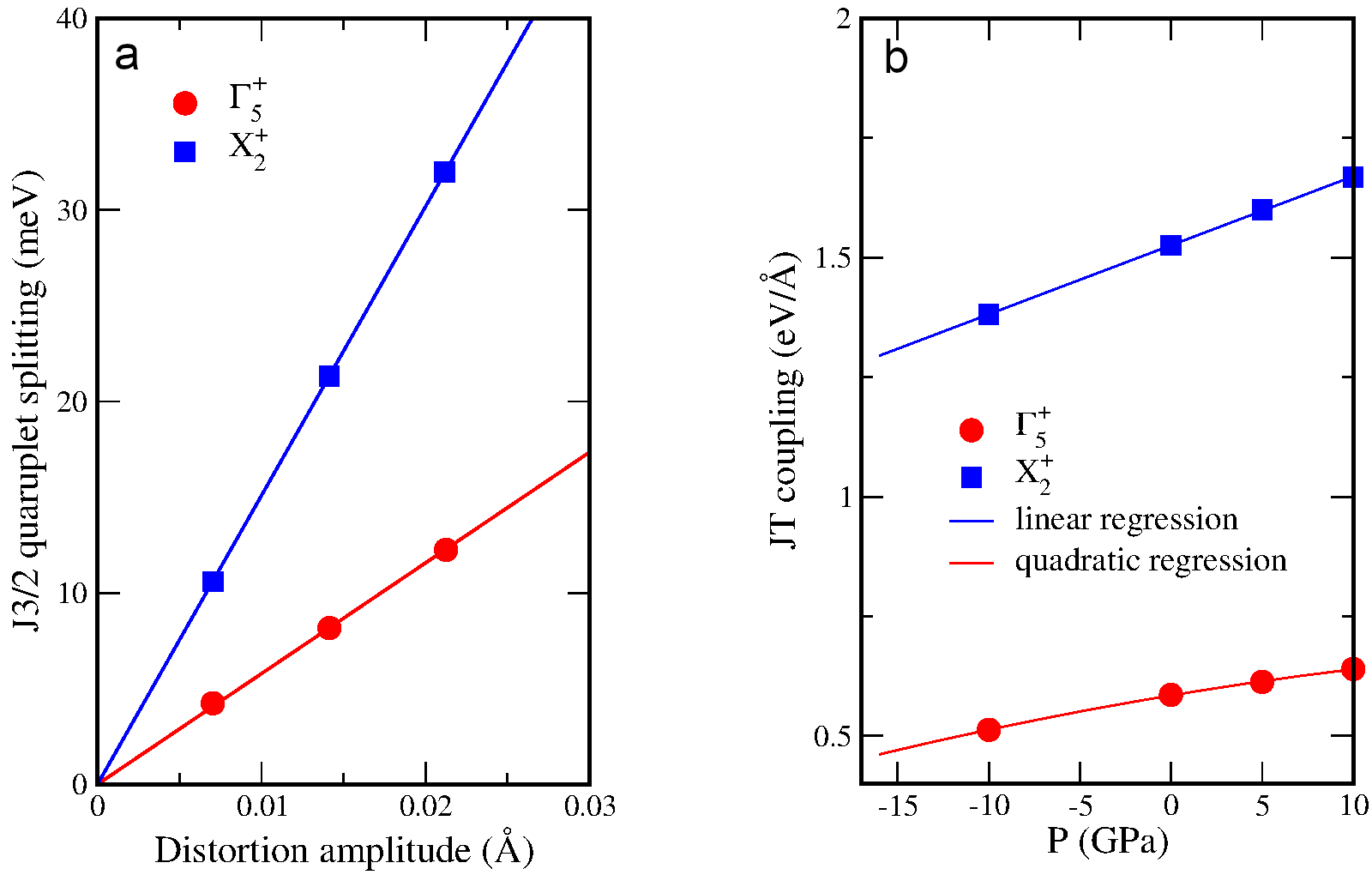}
    \caption{a. Calculated \Jeff quadruplet splitting vs. distortion amplitude for BMRO at ambient pressure is shown by symbols. The lines are linear fits to  the data. b. Calculated JT coupling vs. pressure (symbols). The lines are linear (for $X_2^+$) and quadratic (for $\Gm_5^+$) regression fits to the data. }
    \label{fig:JT}
\end{figure}

Due to the JT coupling $- g_{\Gm} O^{Q}_2 q_{\Gm Q}$, octahedral distortions split the \Jeff quadruplet into two Kramers doublets. The eigenvalues of normalized \Jeff quadrupolar operators $O_2^Q$ are $\pm 1/2$. Hence, the JT splitting of the quadruplet is simply given by $g_{\Gm}q_{\Gm Q}$. As shown in Suppl. Fig.~\ref{fig:JT}a, the calculated  splitting indeed shows a perfect linear scaling vs. the distortion amplitude $q_{\Gm Q}$ within the studied range, hence, no quadratic JT coupling\cite{Iwahara2023} needs to be introduced.
The larger slope in the case of $X_2^+$ distortion is due to a stronger JT coupling to the $E_g$ octahedral distortions.

We repeated the same calculations for a set of volumes corresponding to compressed and expanded BMRO. The resulting JT coupling are depicted in Suppl. Fig.~\ref{fig:JT}b vs. pressure (calculated from volume using our theoretical equation of states, see Sec.~\ref{sec:EoS}); the JT coupling shows a smooth increase vs. pressure allowing to fit it with a simple linear or quadratic regression. As one sees, the $E_g$ JT coupling is almost 3 times stronger than the $T_{2g}$ one. Our values for the JT coupling  calculated by DFT+HI agree well with the DFT results of Ref.~\onlinecite{soh2023spectroscopic}.

\subsection{Calculation $H_{EC}$}
\label{sec:iec}

To determine the effective Hamiltonian $H_{EC}$, we utilize the projective phonon approach explained in detail in the subsequent section. For the phonon calculations in \bmro we adopted the DFPT approach as incorporated in the Vienna Ab initio Simulation Package (VASP)~\cite{PhysRevB.47.558, PhysRevB.54.11169}. These calculations involve the use of DFT with LDA exchange-correlation potential, employing a $2 \times 2 \times 2$ k-mesh and a high energy cutoff of 600 eV together with a total energy convergence of 10$^{-6}$ eV. The choice of an LDA potential has been based on difficulties in converging accurate phonons within the Perdew-Burke-Ernzerhof exchange correlation functional. We note, however, that using LDA in this context has proven to be accurate in determining both volumetric,  bond-length  and pressure properties (See Suppl. Table~\ref{Tab:eos} and Table~\ref{Tab:bondlengths}).
The calculations are based on a $2 \times 2 \times 2$ supercell structure containing 32 f.u. This supercell was generated through the phonopy package~\cite{phonopy-phono3py-JPCM, phonopy-phono3py-JPSJ} starting from a fully relaxed structure with lattice parameter of 7.9992 \AA. 

For the structural relaxation of the smaller 8 f.u. supercell  we used the LDA potential together with a $12 \times 12 \times 12$ k-mesh, an energy cutoff of 500 eV, and a convergence criterion of 10$^{-5}$ eV on the electronic self-consistent cycle. The volume minimization and atomic positions were optimized using a quasi-Newton algorithm with a step width of 0.2~\AA~and a convergence of ionic relaxation of 10$^{-3}$ eV. Detailed information about the optimized structure is available in the cif file, which is included at the conclusion of the Supplementary Material.

\clearpage
\section{Projective Phonon Formalism}

The convention used in phonopy for the dynamical matrix calculation is the following: 
\begin{equation}
    D_{\alpha \beta} (ij, \mathbf{q}) = \frac{1}{\sqrt{m_i m_j}} \sum_{l} \Phi_{\alpha \beta} (i0, jl) exp( i\mathbf{q} \cdot [\mathbf{r}(jl) - \mathbf{r}(i0)]) \ ,
    \label{eq:sup1}
\end{equation}
where 
\begin{equation}
    \Phi_{\alpha \beta} (i0, jl) = \frac{\partial^2 V}{\partial r_{\alpha 0} (i) r_{\beta l} (j) }
\end{equation}
is the corresponding force constant matrix. Here ($\alpha$, $\beta$) represent the Cartesian coordinates, ($i$,$j$) the atomic sites and ($0$,$l$) the unit-cell index. From here on, for simplicity, we will drop the unit-cell indexes. In order to project onto the ligand ion components with specific irreducible representation  coordinates, i.e. $\Phi_{\alpha \beta} (ij) \rightarrow \Phi_{\Gamma \Gamma'}^{QQ'} (ij)$ we apply a projection operator $C_{\Gamma Q}^{\alpha} $ such that $r_{\Gamma Q} = \sum_\alpha C_{\Gamma Q}^{\alpha} r_\alpha$. 
\begin{equation}
    \Phi_{\Gamma \Gamma'}^{QQ'} (ij) = \frac{\partial^2 V}{\partial r_{\Gamma Q} (i) r_{\Gamma Q'} (j)} =  \sum_{\alpha, \beta} (C_{\Gamma Q}^{\alpha})^t \Phi_{\alpha \beta} (ij)(C_{\Gamma ' Q'}^{\beta}) \ .
\end{equation}
Here $\Gamma \in \{T_{2g}; E_g \}$ while $Q$ runs over the corresponding  distortions $ \{ xy, xz, yz; z^2 x^2-y^2 \}$. The Jahn-Teller irreducible representations are constituted from linear combinations of the Cartesian displacements $X_i$, $Y_i$, and $Z_i$ of the ligand ions situated at the vertices of the distorted octahedron. These displacements are measured relative to their positions in an undistorted reference structure, indicated by $x_i^{(0)}, y_i^{(0)}, z_i^{(0)}$. Employing $X_i$, $Y_i$, and $Z_i$, one can construct JT normal coordinates $Q_i$, which adhere to the irreducible representations of the octahedral symmetry group $O_h$~\cite{bersuker_jahn-teller_2006}.

\begin{table}[ht]
\centering
\renewcommand{\arraystretch}{1.4}
\begin{tabular}{c | c c}
\toprule
$A_{1g}$                  & $Q_1$      & $(X_2-X_5+Y_3-Y_6+Z_1-Z_4)/\sqrt{6}$    \\\\
\hline
{$E_g$}      & $Q_2$      & $(X_2-X_5-Y_3+Y_6)/2$                   \\
                          & $Q_3$      & $(2Z_1-2Z_4-X_2+X_5-Y_3+Y_6)/2\sqrt{3}$ \\\\
\hline
{$T_{2g}$}   & $Q_{yz}$   & $(Z_3-Z_6+Y_1-Y_4)/2$                   \\
                          & $Q_{xz}$   & $(X_1-X_4+Z_2-Z_5)/2$                   \\
                          & $Q_{xy}$   & $(Y_2-Y_5+X_3-X_6)/2$                   \\
\hline 
\hline
\end{tabular}%
\caption{The octahedral JT coordinates $Q_i$ are defined based on the Cartesian distortions $X_i$, $Y_i$, and $Z_i$ of the system.}
\label{Tab:JT_definition}
\end{table}

We test the consistency of the method in two cases: 
\begin{enumerate}
    \item By comparing the $\Gamma-$point projected frequencies against the experimental and computed ones of a non-magnetic double perovskite containing Re, i.e. Ba$_2$WReO$_6$.
    \item By comparing the projected-phonon spectra to the fully computed one for \bmro.
\end{enumerate}

The implementation of the projective phonon formalism for calculation of EC is available on GitHub at the following link~\hyperlink{https://github.com/dariofiosca/ElastiCouplings.git}{https://github.com/dariofiosca/ElastiCouplings.git}

\clearpage
\subsection{Comparison with Ba$_2$WReO$_6$}

We start by addressing point (1).  Ba$2$WReO$6$ is a non-magnetic double perovskite whose $\Gamma$-point irreducible phonon frequencies were determined experimentally through Raman spectroscopy~\cite{PASZTOROVA2023124184} and via theoretical first-principles involving both DFT and lattice dynamic calculations (LDC)~\cite{Diao, Rodrigues}. The $\Gamma$-point frequencies can be directly derived from equation~\ref{eq:sup1} where $D_{\Gamma \Gamma'}^{QQ'} (0) = 1/\sqrt{m_1 m_j} \sum_{l} \Phi_{\Gamma \Gamma'}^{QQ'} (0, l)$, and their comparison is detailed in Suppl. Table~\ref{Tab:bwro}.

The phonon spectra alongside on-site and EC were calculated subsequent to a thorough DFT-based relaxation of the volume and internal atomic coordinates, using the LDA potential. For the relaxation process, a $10 \times 10 \times 10$ k-mesh was utilized, a 400 eV cutoff for the plane wave basis, and an electronic self-consistent cycle convergence of 10$^{-4}$ eV. The final lattice parameters are listed in Table~\ref{Tab:bwro}.

\noindent For phonon calculations, a $2 \times 2 \times 2$ supercell was constructed using the phonopy tool, and the DFPT routine in VASP was employed. The calculations were performed with an enhanced energy convergence of 10$^{-6}$ eV and an increased energy cutoff of 600 eV. Calculations included both on-site, nearest-neighbor EC, and next-nearest-neighbor EC.

\begin{table}[!h]
  	\caption{\label{Tab:bwro}  
  		%\justifying
  		 Comparison between lattice constants for Ba$_2$WReO$_6$. The $\dagger$ is used to mark the result of this work. 
  	}
   			\renewcommand{\arraystretch}{1.4}
\begin{tabular}{ c | c c c c c c c c }
\hline
\hline
     & DFT $\dagger$ & Exp. 100 K~\cite{PASZTOROVA2023124184} &Exp. 300 K~\cite{PASZTOROVA2023124184} \\
\hline
$a$ (\AA)    & 8.1046  & 8.1010 & 8.1120 \\
\hline
\hline
\end{tabular}
\label{Tab:bwro}
\end{table}

\begin{table}[!h]
  	\caption{\label{Tab:bondlengths}  
  		%\justifying
  		 The frist column represents the computed irreducible frequencies as $\Gamma$ obtained from phonopy, while the second are the ones obtained from the projective method including onsite JT and both nearest and next nearest neighbors EC. The $\dagger$ represents the results of this work. The third colum includes the experimentally defined Raman modes at 300 K and the comparison with experimental data at 300 K on the ceramic material, together with two theoretical calculations of phonon spectra.
  	}
   			\renewcommand{\arraystretch}{1.4}
\begin{tabular}{ c | c c c c c c c }
\hline
\hline
R. mod.    & Phonopy$\dagger$ & Projected$\dagger$ & Exp. Ref~\cite{PASZTOROVA2023124184} & Exp Ref.~\cite{Diao} & DFT Ref.~\cite{Diao} & LDC Ref.~\cite{Rodrigues}    \\
\hline
$E_g$    & 534  & 492 & 536 & 538 & 614 & 522 \\
$T_{2g}$ & 421  & 418 & 439 & 441 & 405 & 432 \\
$A_{1g}$ & 780  & 784 & 807 & 812 & 833 & 776 \\
\hline
\hline
\end{tabular}
\label{Tab:bwro}
\end{table}

\subsection{Comparison phonons in BMRO}

From the projected force constant matrix $\Phi_{\Gamma \Gamma'}^{QQ'}(ij)$ it is possible to recalculate the Dynamical Matrix 
\begin{equation}
    D_{\Gamma \Gamma'}^{QQ'} (\mathbf{q}) = \frac{1}{\sqrt{m_Q m_{Q'}}} \sum_{ij} \Phi_{\Gamma \Gamma'}^{QQ'} (ij) exp (i \mathbf{q} \cdot [ \mathbf{r}_{\Gamma}^{Q}(i) - \mathbf{r}_{\Gamma'}^{Q'}(j)])
\end{equation}
where ($m_{Q}$, $m_{Q'}$) are the oxygen masses, and $\mathbf{q}$ the wave vector. The equation of motion can be derived to be
\begin{equation}
    \sum_{\Gamma' \gamma'} D_{\Gamma \Gamma'}^{QQ'} (\mathbf{q}) e_{\Gamma'}^{Q'} (\mathbf{q} \nu) = [\omega(\mathbf{q} \nu)]^2 e_{\Gamma}^{Q} (\mathbf{q} \nu)
\end{equation}
and via diagonalization of the equation of motion one obtains the phonon spectra.

This methodology serves to evaluate the precision of the projective phonon technique in comparison to the full phonon calculation of the 32 f.u. supercell. It's important to highlight that the inclusion of next-nearest neighbors (NNN) EC is crucial for achieving a high degree of accuracy in the phonon band structure at the $\Gamma$ and $X$ points. 

While incorporating ECs beyond NNN can refine the phonon spectra comparison, this does not necessarily yield a physically comprehensive solution, as the octahedra beyond NNN are impacted by the supercell's periodic boundary conditions.

\begin{figure}[!h]
    \centering
    \includegraphics[scale=0.7]{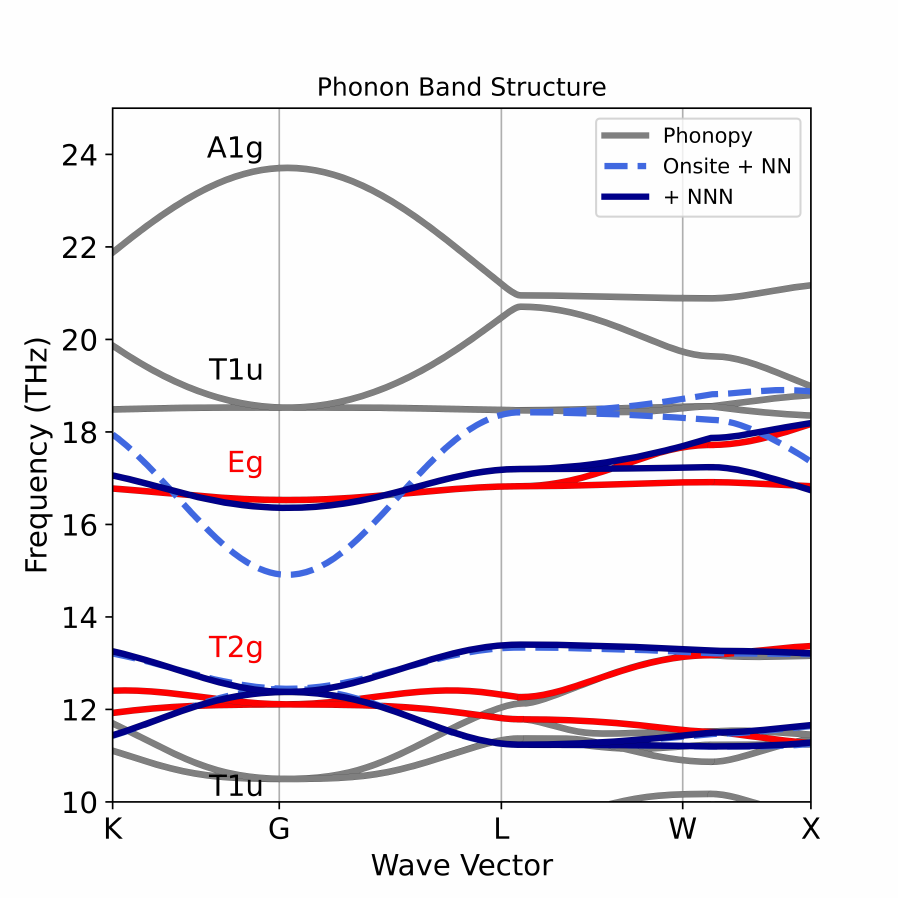}
    \caption{Comparison between primitive cell phonons of \bmro computed with phonopy and the reconstructed phonons reconstructed from the projected force constant matrix $\Phi_{\Gamma \Gamma'}^{QQ'} (ij)$. The red lines mark the $E_g$ and $T_{2g}$ phononic bands, while the dotted light blue are the reconstructed bands from onsite + NN EC and the solid blue line mark the reconstructed bands from onsite + NN + NNN.}
    \label{fig:phonons}
\end{figure}

We note that the current has its main limitation in the DFPT implementation of VASP, which only calculates the force constant at the $\Gamma$-point. An accurate computation of the force constants at other $\mathbf{q}$ points  requires the use of significantly larger supercell structures. However, scaling up the supercell size poses substantial computational challenges. Fortunately, for BMRO extensive supercell calculations are not mandatory.  As demonstrated in Suppl. Fig.~\ref{fig:phonons}, we achieve satisfactory convergence at the $\Gamma$ and $X$ high-symmetry points by incorporating on-site, nearest-neighbor (NN), and NNN intersite elastic couplings.

\clearpage
\subsection{Phonon Band-structures Ba$_2$WReO$_6$ and Ba$_2$MgReO$_6$}

\begin{figure}[!h]
    \centering
    \includegraphics[scale=0.8]{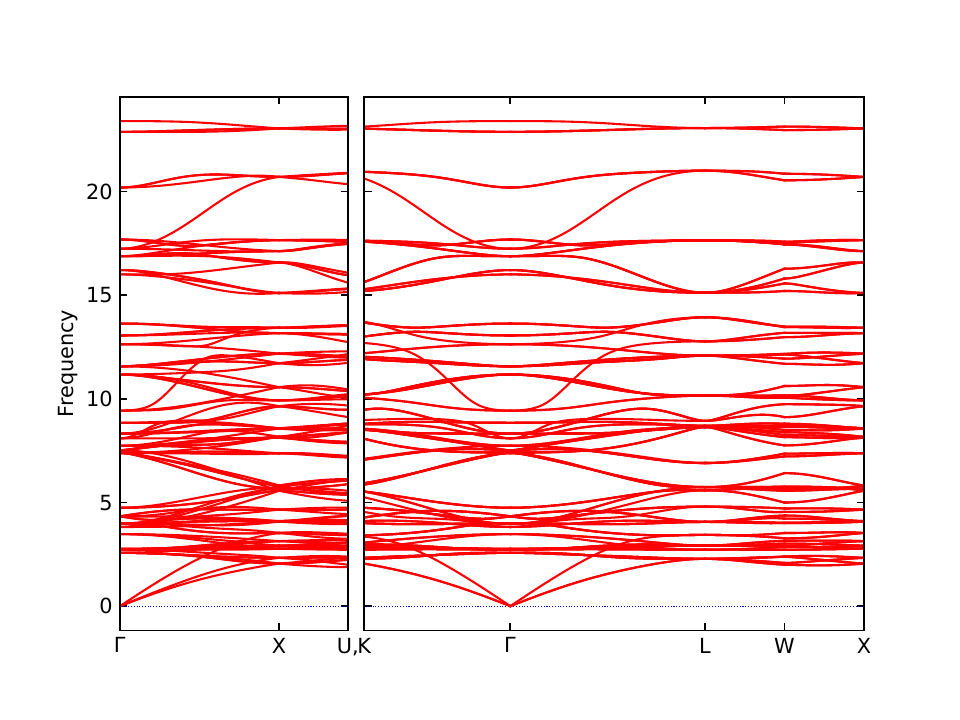}
    \caption{Phonon bandstructure of Ba$_2$WReO$_6$. Frequencies are in THz.}
    \label{fig:}
\end{figure}

\begin{figure}[!h]
    \centering
    \includegraphics[scale=0.8]{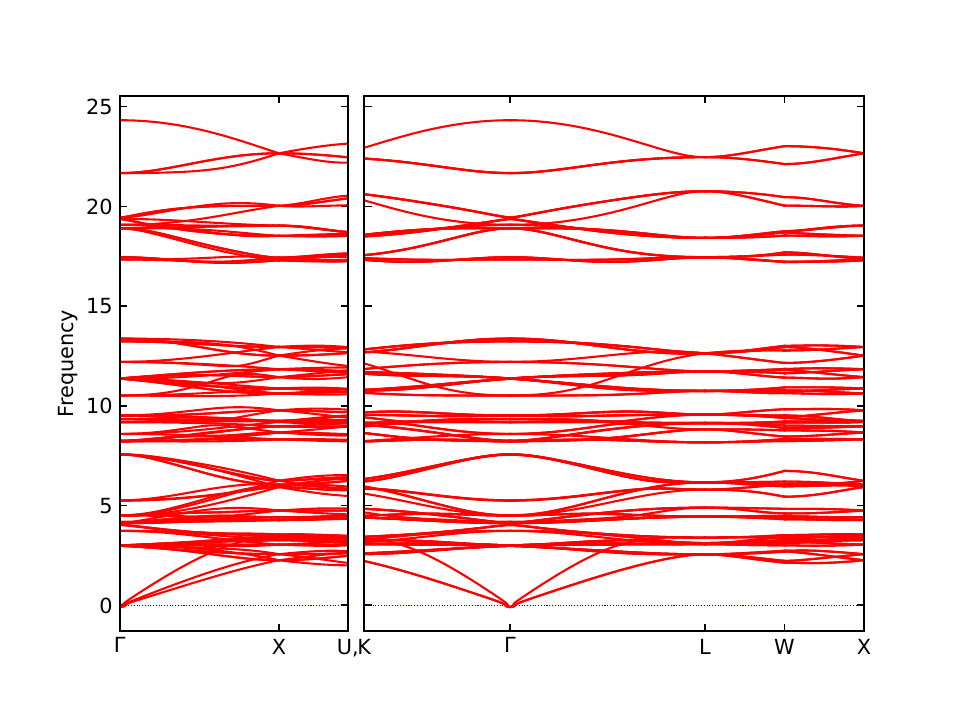}
    \caption{Phonon bandstructure of Ba$_2$MgReO$_6$. Frequencies are in THz.}
    \label{fig:}
\end{figure}

\clearpage
\section{Equation of State} \label{sec:EoS}

To compute the volumes at different pressure we fit the pressure as function of volume with the Birch-Murnaghan Equation of State (EOS)
\begin{equation}
    P = \frac{3}{2} B_0 \Big[ \Big( \frac{V}{V_0} \Big)^{-7/3} -  \Big( \frac{V}{V_0} \Big)^{-5/3} \Big] \times \Big\{ 1 + \frac{3}{4} (B'_0 - 4)\Big[ \Big( \frac{V}{V_0} \Big)^{-2/3} - 1 \Big] \Big\}
\end{equation}

The volumetric calculations were done in DFT by employing the VASP package with the 8 f.u. cubic supercell, starting from the fully relaxed solution described in Section~\ref{sec:iec}. The LDA potential is used, together with a k-mesh of $12 \times 12 \times 12$ and energy cutoff of 600 eV with energy convergence of the self consistent cycle of 10$^{-5}$ eV. 
The results in Table~\ref{Tab:eos} have been compared with both experimental and first-principles calculation from Ref.~\cite{} and Ref.~\cite{DAR2018137} respectively.

\begin{table}[!h]
  	\caption{\label{Tab:eos}  
  		%\justifying
  		 Calculated bulk modulus ($B_0$ in GPa) and derivative of bulk modulus ($B'_0$).
  	}
   			\renewcommand{\arraystretch}{1.4}
\begin{tabular}{ c c c }
\hline
\hline
        & DFT (this work) & DFT~\cite{DAR2018137}    \\
\hline
$B_0$ (GPa)    & 197.47      & 158.27 \\
$B'_0$         & 4.51        & 4.80   \\
\hline
\hline
\end{tabular}
\end{table}

\begin{table}[!h]
  	\caption{\label{Tab:bondlengths}  
  		%\justifying
  		 Calculated lattice constant and bondlength distances.
  	}
   			\renewcommand{\arraystretch}{1.4}
\begin{tabular}{ c c c c }
\hline
\hline
    & DFT (this work) & Exp. Ref~\cite{Hirai2020_PRR}  & DFT~\cite{DAR2018137}    \\
\hline
$a$ (\AA)    & 7.99292         & 8.0802 & 8.1790 \\
Re-O (\AA) & 1.92            & 1.96   & 1.95   \\
Mg-O (\AA) & 2.08            & 2.08   & 2.14   \\
Ba-O (\AA) & 2.83            & 2.85   & 2.89  \\
\hline
\hline
\end{tabular}
\end{table}

\begin{figure}[!h]
    \centering
    \includegraphics[scale=0.4]{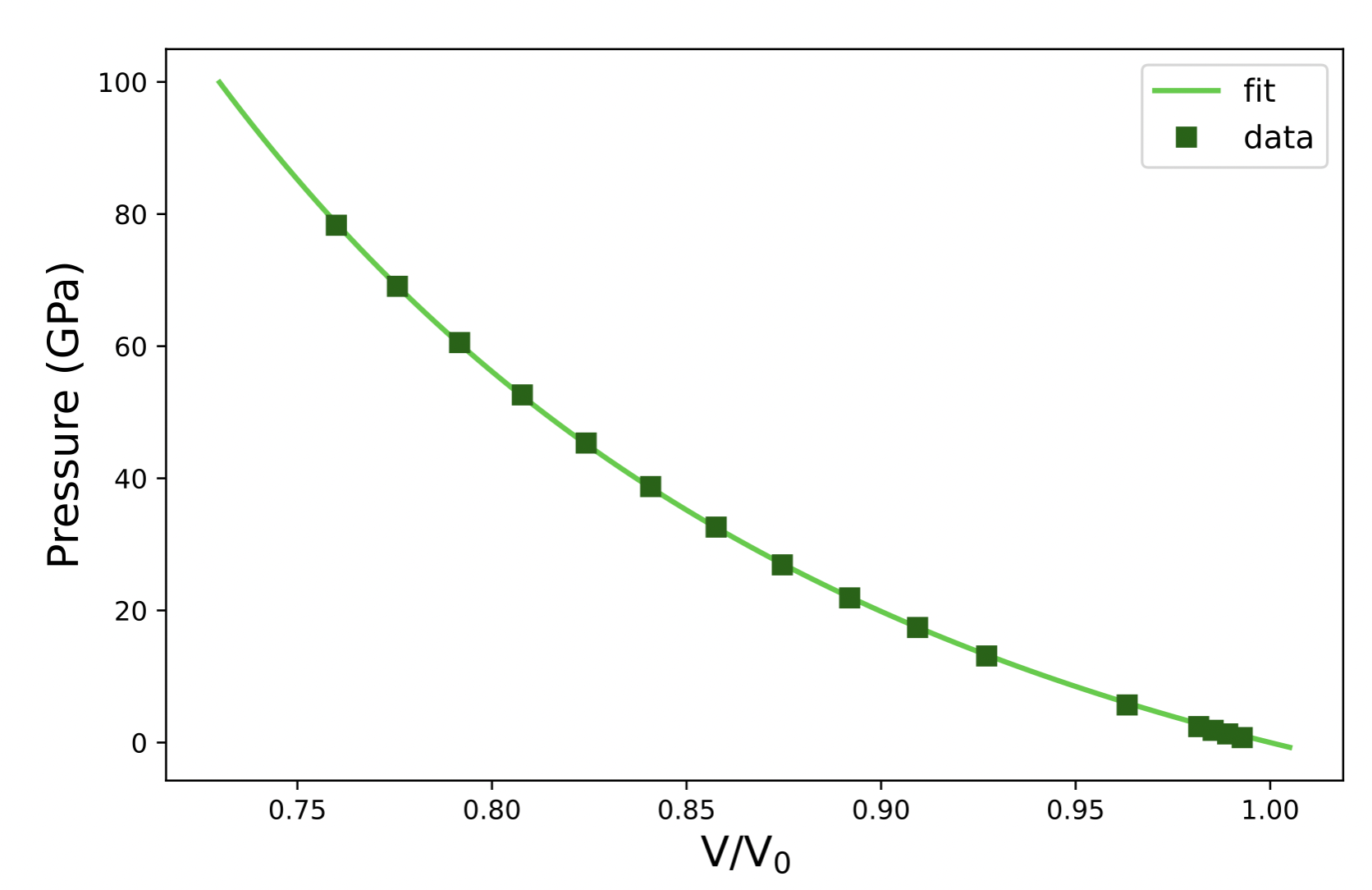}
    \caption{Fit of Pressure vs volumetric DFT solution with Birch-Murnaghan EOS.}
    \label{fig:eos}
\end{figure}

From the EOS we derived for both the Experimental and DFT-relaxed structures, the evolution of the lattice constant with pressure. In the following table we report the values used to calculate $H_{IEI}$ and $H_{IEC}$. 

\begin{table}[!h]
  	\caption{\label{Tab:pressure}  
  		%\justifying
  		 Calculated lattice constant from \bmro EOS employed for the calculation of IEI and EC.
  	}
   			\renewcommand{\arraystretch}{1.4}
\begin{tabular}{ c c c}
\hline
\hline
Pressure (GPa)    & IEI $a$ (\AA) & IEC $a$ (\AA)\\
\hline
-10   & 8.3058   & 8.22258\\
-5    & 8.1810   & 8.09898\\
 0    & 8.0802   & 7.99292\\
 5    & 7.9954   & 7.91524\\
10    & 7.9221   & 7.84270\\
\hline
\hline
\end{tabular}
\end{table}

\section{Intersite Exchange Interactions}

\begin{figure}[!h]
    \centering
    \includegraphics[scale=0.8]{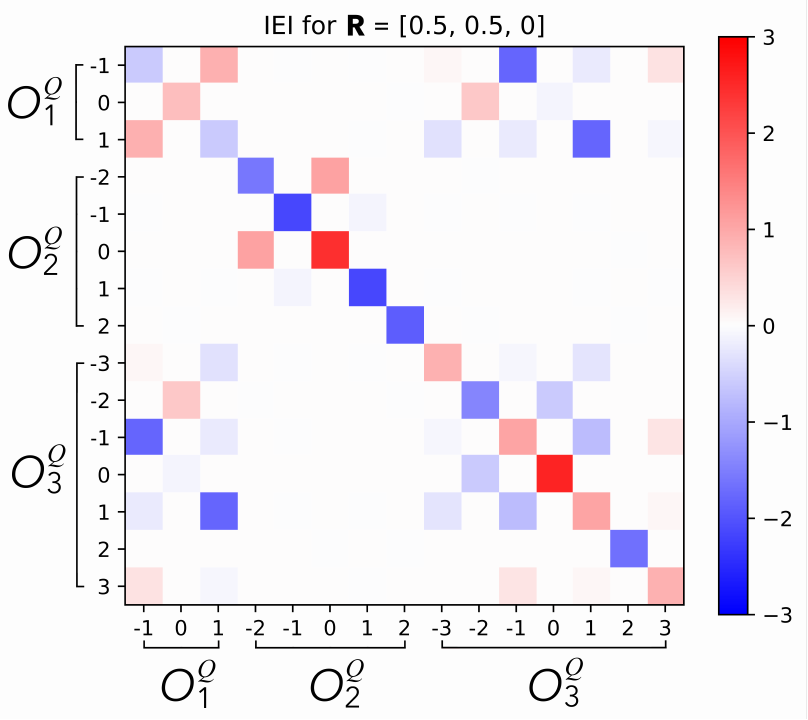}
    \caption{Color map of the IEI $V_{KK'}^{QQ'}$ in BMRO at pressure 0 GPa for the [1/2,1/2,0] Re-Re pair. All values are in meV. The numerical list of $V_{KK'}^{QQ'}$ is given in Suppl. Table~\ref{Tab:SEI}.}
    \label{fig:bmro_iei}
\end{figure}

In Suppl. Fig.~\ref{fig:bmro_iei} we plot the IEI in BMRO at zero pressure as a color map; the IEI values  are also listed in  Suppl. Table~\ref{Tab:SEI} for several values of pressure.  One sees that the quadrupole IEI block includes, apart from antiferroic $e_g$ coupling, also  ferroic $t_{2g}$ ones. These $t_{2g}$ ferroic IEI have not been previously included in  model electronic effective Hamiltonians of $d^1$ DP~\cite{Chen2010,Svoboda2021,Iwahara2023,Kubo2023}. The $t_{2g}$ IEI  are the reason why the F\xy quarupole order dominates over the AF\xs one when the EL coupling is neglected. 

One may also notice significant octupolar IEI, which are the largest couplings among time-odd IEI.

  	\renewcommand\floatpagefraction{0.1}
  \begin{table}[!h]
  	\caption{\label{Tab:SEI}  
  		%\justifying
  		 Calculated IEI $V_{KK'}^{QQ'}$ for the $J_{eff}$=3/2  multiplet. First two columns list
  		$Q$ and $Q'$, respectively. Third and fourth column displays the $KQ$ and $K'Q'$ tensors in the  Cartesian representation, respectively. The last three columns displays the values of IEI (in meV) for the [1/2,1/2,0] nearest-neighbor Re-Re bond in \bmro  for pressure 10~GPa, ambient condition and -16 GPa, respectively.
  	}
	\begin{center}
		\begin{ruledtabular}
			\renewcommand{\arraystretch}{1.2}
			\begin{tabular}{c c c c c c c}
\multicolumn{4}{c}{Pressure} & 10 GPa & Ambient & -16 GPa \\
\multicolumn{6}{c}{ } & (Ba$_2$CaReO$_6$ volume) \\
		\hline
\multicolumn{7}{c}{Dipole-Dipole} \\
 				\hline
-1 & -1   & y  & y  &   -0.79  &   -0.59  &   -0.31  \\
 0 &  0   & z  & z  &    1.00  &    0.73  &    0.38  \\
 1 & -1   & x  & y  &    0.99  &    0.91  &    0.73  \\
 1 &  1   & x  & x  &   -0.79  &   -0.59  &   -0.31  \\
		\hline
		\hline
\multicolumn{7}{c}{Quadrupole-Quadrupole} \\
\hline
-2 & -2   & xy  & xy  &   -2.07  &   -1.59  &   -0.94  \\
-1 & -1   & yz  & yz  &   -2.67  &   -2.14  &   -1.38  \\
 0 & -2   & z$^2$  & xy  &    1.19  &    1.08  &    0.84  \\
 0 &  0   & z$^2$  & z$^2$  &    3.05  &    2.43  &    1.55  \\
 1 & -1   & xz  & yz  &   -0.10  &   -0.10  &   -0.08  \\
 1 &  1   & xz  & xz  &   -2.67  &   -2.14  &   -1.38  \\
 2 &  2   & x$^2$-y$^2$  & x$^2$-y$^2$  &   -2.36  &   -1.89  &   -1.23  \\
\hline
		\hline
\multicolumn{7}{c}{Octupole-Octupole} \\
\hline
-3 & -3   & y(x$^2$-3y$^2$)  & y(x$^2$-3y$^2$)  &    1.16  &    0.88  &    0.50  \\
-2 & -2   & xyz  & xyz  &   -1.85  &   -1.42  &   -0.84  \\
-1 & -3   & yz$^2$  & y(x$^2$-3y$^2$)  &   -0.08  &   -0.09  &   -0.08  \\
-1 & -1   & yz$^2$  & yz$^2$  &    1.23  &    1.03  &    0.75  \\
 0 & -2   & z$^3$  & xyz  &   -0.64  &   -0.60  &   -0.49  \\
 0 &  0   & z$^3$  & z$^3$  &    3.13  &    2.56  &    1.75  \\
 1 & -3   & xz$^2$  & y(x$^2$-3y$^2$)  &   -0.32  &   -0.28  &   -0.21  \\
 1 & -1   & xz$^2$  & yz$^2$  &   -0.83  &   -0.76  &   -0.61  \\
 1 &  1   & xz$^2$  & xz$^2$  &    1.23  &    1.03  &    0.75  \\
 2 &  2   & z(x$^2$-y$^2$)  & z(x$^2$-y$^2$)  &   -2.06  &   -1.65  &   -1.06  \\
 3 & -1   & x(3x$^2$-y$^2$)  & yz$^2$  &    0.32  &    0.28  &    0.21  \\
 3 &  1   & x(3x$^2$-y$^2$)  & xz$^2$  &    0.08  &    0.09  &    0.08  \\
 3 &  3   & x(3x$^2$-y$^2$)  & x(3x$^2$-y$^2$)  &    1.16  &    0.88  &    0.50  \\
		\hline
		\hline
\multicolumn{7}{c}{Dipole-Octupole} \\
\hline
-1 & -3   & y  & y(x$^2$-3y$^2$)  &    0.07  &    0.08  &    0.07  \\
-1 & -1   & y  & yz$^2$  &   -2.25  &   -1.79  &   -1.15  \\
-1 &  1   & y  & xz$^2$  &   -0.23  &   -0.21  &   -0.17  \\
-1 &  3   & y  & x(3x$^2$-y$^2$)  &    0.36  &    0.31  &    0.23  \\
 0 & -2   & z  & xyz  &    0.69  &    0.61  &    0.47  \\
 0 &  0   & z  & z$^3$  &   -0.15  &   -0.12  &   -0.09  \\
 1 & -3   & x  & y(x$^2$-3y$^2$)  &   -0.36  &   -0.31  &   -0.23  \\
 1 & -1   & x  & yz$^2$  &   -0.23  &   -0.21  &   -0.17  \\
 1 &  1   & x  & xz$^2$  &   -2.25  &   -1.79  &   -1.15  \\
 1 &  3   & x  & x(3x$^2$-y$^2$)  &   -0.07  &   -0.08  &   -0.07  \\
  			\end{tabular}
\end{ruledtabular}
\end{center}
\end{table}

\newpage

\section{Symmetry analysis of IEI matrix}

The form of intersite interaction matrix $V_{KK'}^{QQ'}$ can be understood on the basis of symmetry arguments only. 

To that end one needs first to classify the multipolar operators $O_K^Q$ among irreducible representations of the $O_h$ cubic point group.  As discussed previously in the literature~\cite{abragam_bleaney_book,Shiina1997,Santini2009}, by reduction of the direct product $\Gamma_8\times \Gamma_8$ one finds the following multipoles IREPs to be realizable within a $\Gamma_8$ multiplet: monopole $A_g^+$, dipoles $T_{1g}^{d-}$, quadrupoles ($E_g^{+}$, $T_{2g}^+$) and octupoles ($A_{1u}^-$, $T_{1u}^{o-}$, and $T_{2u}^-$). Here the subscript $g(u)$ labels even (odd) behavior under inversion of the corresponding Cartesian tensor, the subscript $+(-)$ labels even(odd) operators under time reversal. Multipolar operators acting within a single atomic shell (Re 5$d$) are parity-even; in this case the $g(u)$ Cartesian tensors correspond to charge (magnetic) multipoles\cite{Santini2009,Kuramoto2009}. Hence, we drop the superfluous superscripts $+$ and  $-$ from now on.  The superscripts $d$ and $o$ are needed to distinguish the $T_{1u}$ dipoles and octupoles.

\begin{figure}[!h]
    \centering
    \includegraphics[scale=0.8]{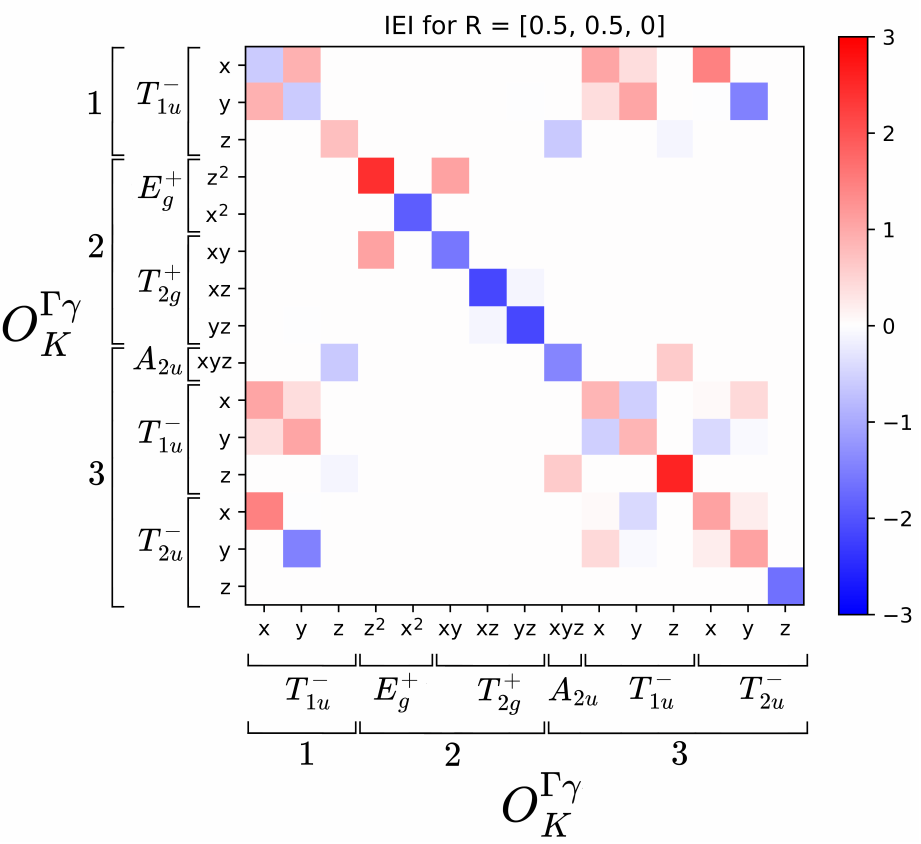}
    \caption{The same BMRO IEI matrix $\hat{V}$ as shown in Suppl. Fig.~\ref{fig:bmro_iei} replotted using operators $O_K^{\Gamma\gamma}$ that form the bases of cubic IREPs $\Gamma$.}
    \label{fig:bmro_iei_ireps}
\end{figure}

The dipole ($O_1^Q$) and quadrupole ($O_2^Q$) operators  employed in Eq. 2 of the main text already form proper IREP bases for the dipoles and quadrupoles. In the case of octupoles, a basis rotation specified, e.~g., in Ref.~\cite{Shiina1997} is applied to transform the $O_3^Q$ tensors  into the bases of $T_{1u}^{o}$, and $T_{2u}$ IREPs. In Suppl. Fig.~\ref{fig:bmro_iei_ireps} we plot the IEI matrix $\hat{V}$ in the operators  $O_K^{\Gamma\gamma}$, which is obtained from that in Suppl. Fig.~\ref{fig:bmro_iei} after this basis transformation. $\Gamma$ and $\gamma$ label in the operator $O_K^{\Gamma\gamma}$ the IREP and projection, respectively.  One sees that the time-even and time-odd multipoles do not couple, as expected, with the IEI matrix separating into dipole-octupole and quadrupole-quadrupole blocks. 

In order to explain the structure of those blocks we consider two Re ions connected by the [1/2,1/2,0] lattice vector. The resulting two-site cluster is invariant under the orthorhombic $D_{2h}$ subgroup of $O_h$, since it is invariant under the inversion, a two-fold rotation about the directions of the bond as well as two-fold rotations about the in-plane ([1-10]) and out-of-plane ([001]) directions, which are orthogonal to the bond.  The corresponding subductions of all $O_h$ IREPs to $D_{2h}$:
\begin{equation}
\mathrm{IREP}_{\Gamma\gamma}(O_h) \to \mathrm{IREP}_{A}(D_{2h}) \oplus \mathrm{IREP}_{B}(D_{2h}) \oplus...
\end{equation}
are given in Ref.~\cite{altmann1994point}, where $A$ and $B$ label various $D_{2h}$ IREPs (all of them are one-dimensional). One should notice that this subduction is accompanied by an IREP basis change corresponding to the coordinate frame rotation by $\pi/4$ about the $z$ axis, i.~e. $x' = (x-y)/\sqrt{2}$ and $y' =(x+y)/\sqrt{2}$, where the prime labels the axes of $D_{2h}$ frame. Non-zero matrix elements in the IEI matrix block $\hat{V}_{KK'}^{\Gamma\Gamma'}$ arise whenever the subductions of two cubic IREPs $\Gamma$ and $\Gamma'$ contain the same $D_{2h}$ IREP.

\vspace{3mm}
\underline{Dipole-dipole block}
\vspace{1mm}

The subduction of $O_h$ IREP $T_{1u}$  to $D_{2h}$ reads
$$
T_{1u}\to B_{1u}\oplus B_{2u} \oplus B_{3u}.
$$

There will be thus the following non-zero terms in the dipole-dipole interaction block ($T_{1u} \times T_{1u}$):
$$
V_{B1}^d O^{B_{1u}}_1(i)O^{B_{1u}}_1(j), \qquad   
V_{B2}^d O^{B_{2u}}_1(i)O^{B_{2u}}_1(j), \qquad  
V_{B3}^d O^{B_{3u}}_1(i)O^{B_{3u}}_1(j). 
$$
where the operator bases of $B_{1u}$, $B_{2u}$, and $B_{3u}$ are $O_1^{T_{1u} z}$, $O_1^{T_{1u} y}$, and $O_1^{T_{1u} x}$, respectively\cite{altmann1994point}.
Taken into account the basis rotation between $O_h$ and $D_{2h}$ coordination frame, one finds that $O_1^{B_{1u}}= O_1^{T_{1u} z}\to O_1^{T_{1u} z}$,  $O_1^{B_{2u}}=O_1^{T_{1u} y}\to (-O_1^{T_{1u} x}+O_1^{T_{1u}y})/\sqrt{2}$, and $O_1^{B_{3u}}=O_1^{T_{1u} x}\to (O_1^{T_{1u} x}+O_1^{T_{1u}y})/\sqrt{2}$ upon the inverse rotation to the original cubic frame. Hence, the interaction matrix in the cubic frame is
$$
\left[
\begin{matrix}
  V_1^d & V_2^d &  \\
  V_2^d & V_1^d &  \\
   &  & V_{B1}^d
 \end{matrix}
\right],
$$
where $V_1^d=(V_{B2}^d+V_{B3}^d)/2$ and $V_2^d=(V_{B3}^d-V_{B2}^d)/2$. The operators acting within the bond plane ($xy$) have the same diagonal couplings, but for the out-of-plane direction ($z$) it is different. In addition, there is an off-diagonal coupling between the in-plane operators. This form agrees with the dipole-dipole block calculated ab initio (Suppl.~Fig.~\ref{fig:bmro_iei_ireps}).

\vspace{3mm}
\underline{Quadrupole-quadrupole block}
\vspace{1mm}

The subductions of $O_h$ IREPs $E_g$ and $T_{2g}$ to $D_{2h}$ read
\begin{align*}
    E_{g}\to A_g\oplus B_{1g}, \\
    T_{2g} \to A_g\oplus B_{2g} \oplus B_{3g}.
\end{align*}

Hence, for the $E_g-E_g$ subblock one has two non-zero terms:
$$
V_{Ag}^{1} O^{A_g}_2(i)O^{A_g}_2(j), \qquad   
V_{B1}^q O^{B_{1g}}_2(i)O^{B_{1g}}_2(j).
$$
The operator bases of $D_{2h}$ $A_g$ for quadrupoles ($K=2$) are $O_2^0(=O_2^{E_g z^2})$ and $O_2^2(=O_2^{E_g x^2-y^2})$, see Table 31.6 in Ref.~\cite{altmann1994point}. For the $A_g$ IREP entering into the subduction of $E_g$, the relevant operator is $O_2^{E_g z^2}$. In the case of $B_{1g}$, the operator is $O_2^{-2}(=O_2^{T_{2g} xy})$.  
Taking into account the $\pi/4$ rotation one finds  $O_2^{E_g z^2}$ unchanged, while $O_2^{T_{2g} xy}$ transforms into $O_2^{E_{g} x^2-y^2}$.  Hence, in the $E_g$ sublock there will be  two distinct diagonal couplings, $V_{Ag}^{1}$ and $V_{B1}^q$,  for $z^2$ and $x^2-y^2$ quadrupoles, respectively.

 In the case of the $T_{2g}-T_{2g}$ subblock one has three non-zero terms:
$$
V_{Ag}^{2} O^{A_g}_2(i)O^{A_g}_2(j), \qquad   
V_{B2}^q O^{B_{2g}}_2(i)O^{B_{2g}}_2(j) \qquad 
V_{B3}^q O^{B_{3g}}_2(i)O^{B_{3g}}_2(j).
$$
The operator bases of the $A_g$, $B_{2g}$ and $B_{3g}$ IREPs in this case are $O_2^{E_g x^2-y^2}$, $O_2^{T_{2g}yz}$ and $O_2^{T_{2g}xz}$, respectively. One again needs  to take into account the $\pi/4$ rotation that transforms $O_2^{E_g x^2-y^2}\to O_2^{T_{2g} xy}$, $O_2^{T_{2g}yz}\to (-O_2^{T_{2g} xz}+O_2^{T_{2g} yz})/\sqrt{2}$, and  $O_2^{T_{2g}xz}\to (O_2^{T_{2g} yz}+O_2^{T_{2g} xz})/\sqrt{2}$. Hence, one finds the same diagonal coupling $V_1^q=(V_{B2}^{T2g}+V_{B3}^{T2g})/2$ for the $xz$ and $yz$ quadrupoles as well as the off-diagonal coupling $V_2^q=(V_{B3}^{T2g}-V_{B2}^{T2g})/2$ between them; in addition, there is a diagonal coupling $V_{Ag}^{T2g}$ between $xy$ quadrupoles.

Finally, since $A_{1g}$ appears both in the $T_{2g}$ and $E_g$ subduction, it will result in the following coupling
$$
V_{Ag}^{\mathrm{off}}\left[O_2^{E_g z^2}(i)O_2^{E_g x^2-y^2}(j)+(i \leftrightarrow j)\right],
$$ 
which, after the basis rotation, transforms into a coupling between $z^2$ $E_g$ and $xy$ $T_{2g}$.

Therefore, from the derivation above, one obtains the following form for the quadrupole-quadrupole block (in the same operator order as in Suppl.~Fig.~\ref{fig:bmro_iei_ireps}):

$$
\left[
\begin{matrix}
  V_{Ag}^{1} & & V_{Ag}^{\mathrm{off}} & & \\
   & V_{B1g}^q & & & \\
  V_{Ag}^{\mathrm{off}} &  &V_{Ag}^{2} & & \\
   &  & & V_1^q & V_2^q \\
   &  & & V_2^q & V_1^q \\
 \end{matrix}
\right],
$$
which, again, agrees with the ab initio result.

\vspace{3mm}
\underline{Octupoles}
\vspace{1mm}

The analysis of interactions involving the octupoles follows the same lines and is based on the corresponding subductions for the $O_h$ octupole IREPs to $D_{2h}$:
\begin{align*}
    A_{2u}\to B_{1u}, \\
    T_{1u} \to B_{1u}\oplus B_{2u} \oplus B_{3u}, \\
    T_{2u} \to A_{u} \oplus B_{2u} \oplus B_{3u}.
\end{align*}

Due to a large number of non-zero matrix elements, a full derivation of the octupole-dipole and octupole-octupole blocks becomes rather tedious and we do not carry it out here.
Some conclusions can be drawn immediately from the form of the subductions. For example, the $xyz$ $A_{2u}$ octupole may interact only with $T_{1u}$ dipoles and octupoles, but not with the  $T_{2u}$ octupoles; indeed, the corresponding  elements in the ab initio matrix (Suppl. Fig.~\ref{fig:bmro_iei_ireps}) are zero. Moreover, the $B_{1u}$ operators of the dipole and octupole $T_{1u}$  IREPs are $O_1^{T_{1u} z}$ and $O_3^{T_{1u} z}$, respectively. These operators are invariant under the $\pi/4$ rotation. Hence,  only the coupling of $A_{2u}$ octupole with those $T_{1u}$ operators can be non-zero; this is indeed the case in the ab initio $\hat{V}$ matrix. 

\clearpage
\section{Intersite Elastic Couplings}

  	\renewcommand\floatpagefraction{0.1}
  \begin{table}[!h]
  	\caption{\label{Tab:IEC}  
  		%\justifying
  		 The Table displays the calculated values of EC $\Phi_{\Gamma \Gamma'}^{QQ'} (ij)$ for $(ij)$ being the on-site $(ii)$ component, the irreducible  ligand components $(ij)$ connecting NN Re-Re octahedral centers with $\mathbf{R} = [1/2, 1/2, 0]$ and the irreducible  ligand components $(ij)$ connecting NNN Re-Re centers with $\mathbf{R} = [1, 0, 0]$. First two columns list
  		$Q$ and $Q'$, respectively. Units are in eV/\AA$^2$.
  	}
	\begin{center}
		\begin{ruledtabular}
			\renewcommand{\arraystretch}{1.3}
			\begin{tabular}{ c c c c c c }
\multicolumn{6}{c}{On site elastic couplings}\\
\hline
    &   xy    &   yz   &     z2    &    xz   &   x2-y2 \\
    &   9.633 & 9.633 &   21.822      & 9.633 & 21.822 \\
\hline
\multicolumn{6}{c}{Nearest-Neighbors EC for $\mathbf{R} = [0.5, 0.5, 0]$}\\
 				\hline
    &   xy    &   yz   &     z2    &    xz   &   x2-y2 \\
    xy   &   0.584   &  0  &  -0.463  &  0    &  0\\
    yz   &   0  &  -0.229  &  0 &   -0.250  &  0\\
    z2   &   -0.463   &  0  &  0.532 &   0   &  0\\
    xz   &   0   &  -0.250  &  0  &   -0.229  &  0\\
  x2-y2  &   0  &  0  &  0  &   0  &  -1.745\\
\hline
\multicolumn{6}{c}{Next-Nearest-Neighbors IEC for $\mathbf{R} = [1, 0, 0]$}\\
\hline
    &   xy    &   yz   &     z2    &    xz   &   x2-y2 \\
    xy   &  -0.041   &  0  &  0  &  0    &  0\\
    yz   &   0  &  0.027  &  0 &   0  &  0\\
    z2   &   0   &  0  &  0.446 &   0   &  -0.082\\
    xz   &   0   &  0  &  0  &   -0.041  &  0\\
  x2-y2  &   0  &  0  &  -0.082  &   0  &  0.541\\
\end{tabular}
\end{ruledtabular}
\end{center}
\end{table}

\section{Mean-field approach to the IEI+EL effective Hamiltonian}

We solve the full many-body effective Hamiltonian (eq.~1 of the main text) using a generalized single-site mean-field theory. Both the IEI (eq. 2 of the main text) and EC (eq.~4 ibid) intersite terms are mean-field decoupled; the resulting single-site JT problem in electronic and elastic mean fields is treated fully quantum-mechanically. This mean-field approach to the cooperative JT effect is described in a number of references~\cite{Englman70,Gehring_1975,Polinger2009}; here we implement it for the most general linear JT coupling and mean-field potential within the \Jeff\ space. For consistency, we outline the derivation of this mean-field scheme and fully specify the solution of the single-site JT problem below.

We label the sites in a given ordered phase by the translational vector $\vl$ labeling magnetic unit cells and the basis (sublattice) vector $\vt$ within the magnetic unit cell.  The intersublattice IEI and EC are
$$
V_{KK'}^{QQ'}(\vt\vt')=\sum_{\langle ij \rangle} \sum_{\vl}V_{KK'}^{QQ'}(ij)\delta \left(\vR_{ij}-\vl - \vt' + \vt\right)
$$
and 
$$
\Phi_{\Gm\Gm'}^{QQ'}(\vt\vt')=\sum_{\langle ij \rangle} \sum_{\vl}\Phi_{\Gm\Gm'}^{QQ'}(ij)\delta \left(\vR_{ij}-\vl - \vt' + \vt\right),
$$
respectively.

The mean-field decoupling of the IEI and EC terms then reads
\begin{equation}\label{eq:H_IEI_MF}
H_{\mathrm{IEI}}^{\mathrm{MF}}=\sum_{\vt KQ}\left(O_{K}^Q(\vt)-\frac{1}{2}\langle O_{K}^{Q}(\vt)\rangle\right)V_{KQ}^{\mathrm{MF}}(\vt),
\end{equation}

\begin{equation}\label{eq:H_IEC_MF}
H_{\mathrm{EC}}^{\mathrm{MF}}=\sum_{\tau\Gm}\sum_{Q\in\Gm}\left(q_{\Gm Q}(\vt)-\frac{1}{2}\langle q_{\Gm Q}(\vt)\rangle\right)\Phi_{\Gm Q}^{\mathrm{MF}}(\vt),
\end{equation} 
where the respective IEI and EC mean fields are 
$$
V_{KQ}^{\mathrm{MF}}(\vt)=\sum_{K'Q'\tau'}V_{KK'}^{QQ'}(\vt\vt')\langle O_{K'}^{Q'}(\vt')\rangle
$$ 
and
$$
\Phi_{\Gm Q}^{\mathrm{MF}}(\vt)=\sum_{\Gm'\tau'}\sum_{Q'\in\Gm'}\Phi_{\Gm\Gm'}^{QQ'}(\vt\vt')\langle q_{\Gm' Q'}(\vt')\rangle.
$$

The single-site JT problem in mean-field at a given site $\vt$ is then given by the single-site Hamiltonian (eq. 3 of the main text) plus the mean-field terms:
\begin{equation}\label{eq:H1s_Pq}
    H_{1s} = \sum_{\Gm}\sum_{Q \in \Gm}\left[ \frac{p^{2}_{Q}}{2M_{\Gm}} + \Phi_{\Gm} \frac{q^2_{\Gm Q} }{2} - g_{\Gm} O^{Q}_2 q_{\Gm Q}\right] +H_{\mathrm{IEI}}^{\mathrm{MF}}+H_{\mathrm{EC}}^{\mathrm{MF}},
\end{equation}
where from now on we omit the $\vt$ label for brievity.

We diagonalize the harmonic oscillator terms in (\ref{eq:H1s_Pq}) and quantize accordingly the displacements $q_{\Gm Q}$ in the JT and EC terms. Omitting the constant $-\frac{1}{2}\langle...\rangle$ terms in (\ref{eq:H_IEI_MF}) and (\ref{eq:H_IEC_MF}), the  resulting Hamiltonian for the single-site JT problem reads
\begin{equation}\label{eq:H1s}
H_{1s} = \sum_{\Gm}\sum_{Q \in \Gm}\left[ \hbar \omega_{\Gm}(\hat{n}_{\Gm Q}+1/2) +  (\Phi_{\Gm Q}^{\mathrm{MF}}-g_{\Gm}O^{Q}_2)A_{\Gm}(a^{\dagger}_{\Gm Q}+a_{\Gm Q})\right]+\sum_{KQ}V_{KQ}^{\mathrm{MF}}O_K^Q,
\end{equation}
where we introduced creation(annihilation) operators $a^{\dagger}_{\Gm Q}$($a_{\Gm Q}$) for the local octahedral vibration mode $\Gm Q$, $\hat{n}_{\Gm Q}=a^{\dagger}_{\Gm Q}a_{\Gm Q}$. The oscillator strength $\hbar \omega_{\Gm}$, where $\omega_{\Gm}=\sqrt{\Phi_{\Gm}/M_{\Gm}}$, calculated with ab initio on-site elastic couplings $\Phi_{\Gm}$ (Table~\ref{Tab:IEC}) is equal to 75.3 (49.8)~meV for the $E_g$ ($T_{2g}$) modes, respectively. The quantized displacements $A_{\Gm}(a^{\dagger}_{\Gm Q}+a_{\Gm Q})=q_{\Gm Q}$, where $A_{\Gm}=\sqrt{\hbar/(2M_{\Gm}\omega_{\Gm})}$ is equal to 4.17(5.12)~pm  for  the $E_g$($T_{2g}$) modes.

In order to solve the Hamiltonian (\ref{eq:H1s}) we introduce a normalized mixed (vibronic) basis $|\phi_i>=|M_i\rangle\times|\vn_i\rangle$,  where  $\langle \phi_i|\phi_j\rangle =\delta_{ij}$, the electronic state 
$|M\rangle\equiv|J_{\mathrm{eff}};M\rangle$ is the eigenstate of $\hat{J}_{\mathrm{eff}}$  with the projection $M$,  
$|\vn\rangle \equiv |n_{E_g z^2},n_{E_g x^2-y^2},n_{T_{2g} xy},n_{T_{2g} xz},n_{T_{2g} yz}\rangle\equiv | \tilde{\vn}_{\Gm Q},n_{\Gm Q}\rangle$ is the local phonon modes occupancies. For convenience, we introduce here another notation, $\tilde{\vn}_{\Gm Q}$, to designate the occupancies of all modes except the $\Gm Q$ one. In this basis, the matrix elements of different operators  in (\ref{eq:H1s}) read
\begin{gather}
    \langle \phi_i| O_K^Q|\phi_j\rangle= \delta_{\vn_i,\vn_j} \left[O_K^Q\right]_{M_iM_j}, \nonumber\\
    \langle \phi_i| \hbar\omega_{\Gm}(\hat{n}_{\Gm Q}+1/2)|\phi_j\rangle=\delta_{i,j}\hbar\omega_{\Gm}(n_{\Gm Q}+1/2), \label{eq:matel}\\
    \langle \phi_i|a^{\dagger}_{\Gm Q}+a_{\Gm Q}|\phi_j\rangle =\delta_{M_i,M_j}\delta_{\tilde{\vn}_{\Gm Q}^i,\tilde{\vn}_{\Gm Q}^j}\left(\delta_{n_{\Gm Q}^i,n_{\Gm Q}^j+1}\sqrt{n_{\Gm Q}^i}+\delta_{n_{\Gm Q}^i+1,n_{\Gm Q}^j}\sqrt{n_{\Gm Q}^j}\right)\nonumber.
\end{gather}

We cut the basis $\{|\phi\rangle\}$ at a chosen maximum  phonon number $N_{\mathrm{ph}}^{\mathrm{MAX}}=\mathrm{MAX}\left[\sum_{\Gm Q}n_{\Gm Q}\right]$.
Once all matrix elements $\langle\phi_i|H_{1s}|\phi_j\rangle$ have been evaluated with the help of (\ref{eq:matel}), we diagonalize $H_{1s}$ and recalculate the electronic moments $\langle O_K^Q\rangle$, displacements $\langle q_{\Gm Q}\rangle$, total and free energy for a given site. Having solved the single-site JT problem for all sites $\vt$ in the structure, we recalculate the exchange (\ref{eq:H_IEI_MF}) and elastic (\ref{eq:H_IEC_MF}) mean fields to close the mean-field cycle.

This mean-field approach is implemented in the framework of "McPhase" package~\cite{Rotter2004} as an in-house module. As initial guesses of the mean-field procedure we employ all  1\vk\  structures realizable within the BMRO single conventional unit cell that contains 4 magnetic Re sites.

\section{Effect of the vibronic basis cutoff and Jahn-Teller regime}

In order to benchmark the convergence of the numerical solution of the JT problem (\ref{eq:H1s}) with respect to the vibronic basis cutoff $N_{\mathrm{ph}}^{\mathrm{MAX}}$, we solved this problem, without the mean-field terms, varying  $N_{\mathrm{ph}}^{\mathrm{MAX}}$. The JT problem in this case reduces to the single-site term (eq. 3 of the main text), which in the quantized form reads
\begin{equation}\label{eq:HJT}
H_{\mathrm{JT}} = \sum_{\Gm}\sum_{Q \in \Gm} \hbar \omega_{\Gm}(\hat{n}_{\Gm Q}+1/2) -g_{\Gm}O^{Q}_2 A_{\Gm}(a^{\dagger}_{\Gm Q}+a_{\Gm Q}).
\end{equation}

We employ the values of  parameters ($g_{\Gm}$, $\omega_{\Gm}$, and $A_{\Gm}$) that are calculated for BMRO at ambient pressure (see the previous section and Suppl. Fig.~\ref{fig:JT}b).

\begin{figure}[!h]
    \centering
    \includegraphics[scale=0.7]{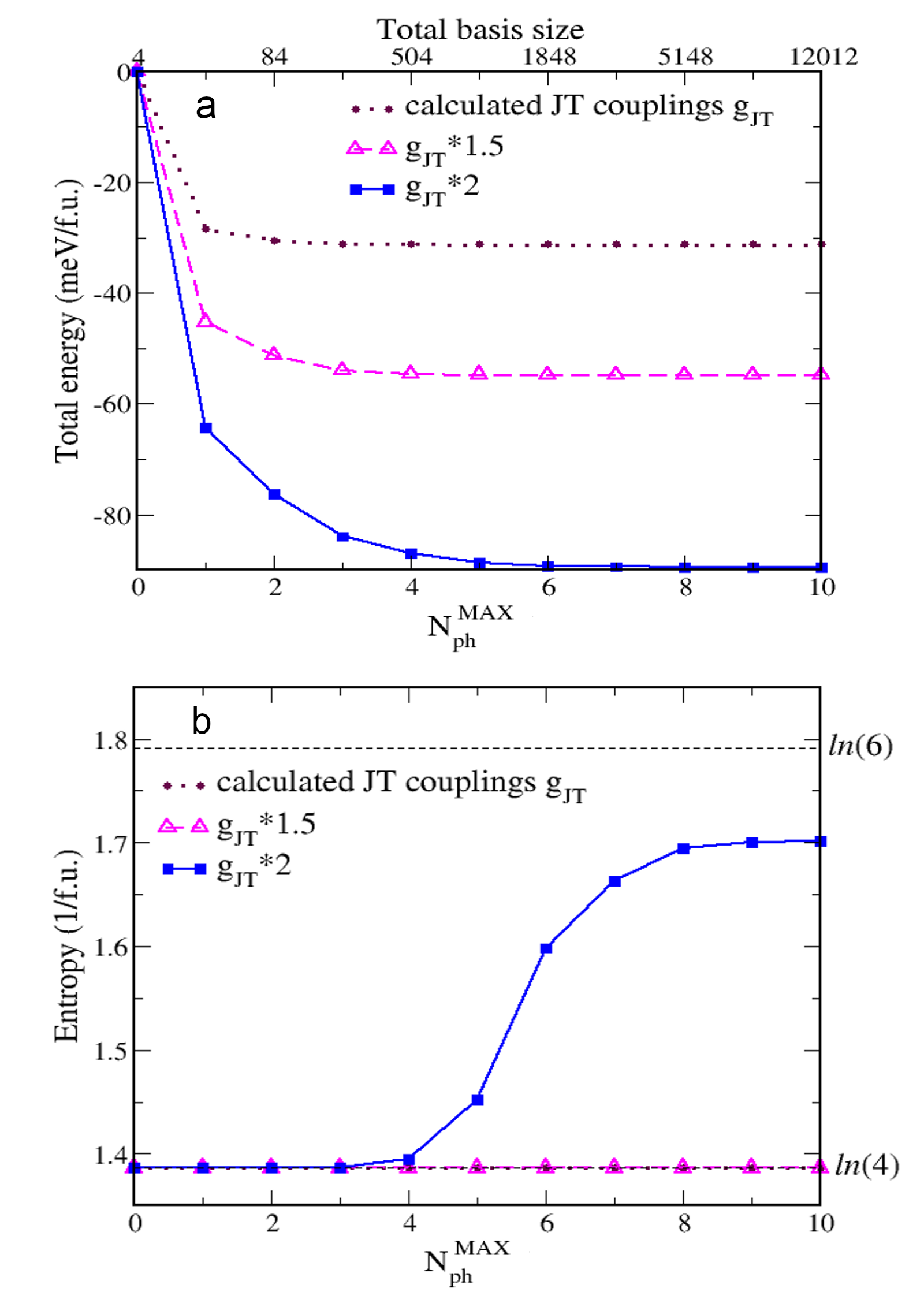}
    \caption{Total energy (a) and entropy (b) of the single-site JT problem ground state as a function of the basis cutoff $N_{\mathrm{ph}}^{\mathrm{MAX}}$. The size of the corresponding basis is indicated by the upper tick labels of panel a.}
    \label{fig:JT_test}
\end{figure}

In Suppl. Figs.~\ref{fig:JT_test}a and \ref{fig:JT_test}b we display the calculated total energy (omitting the zero-point  contribution) and entropy vs. $N_{\mathrm{ph}}^{\mathrm{MAX}}$. One sees that already for $N_{\mathrm{ph}}^{\mathrm{MAX}}=1$, corresponding to the weak coupling treatment of the JT problem, the total energy is quite well converged, while the entropy still corresponds to the ground-state quadruplet and is not affected by including local phonons into the basis. In order to explore the limits of the weak-coupling treatment, we also ran the same calculation  with the JT coupling strength artificially increased. With the JT coupling scaled by the factors 1.5 and 2, the weak-coupling treatment clearly becomes insufficient for the total energy. The ground-state degeneracy markedly  increases once the  JT coupling becomes twice larger, indicating that the JT ground-state solutions are now localized within the "trough" of  the JT potential energy surface~\cite{bersuker_vibronic_1989}.

%\section{Tetragonal crystal field in non hydrostatic BMRO}

\clearpage

\section{Relaxed BMRO structure}

\small
\begin{verbatim}
# CIF file created by FINDSYM, version 7.1.3

data_findsym-output
_audit_creation_method FINDSYM

_cell_length_a     7.9991900000
_cell_length_b     7.9991900000
_cell_length_c     7.9991900000
_cell_angle_alpha  90.0000000000
_cell_angle_beta   90.0000000000
_cell_angle_gamma  90.0000000000
_cell_volume       511.8444957459

_symmetry_space_group_name_H-M "F 4/m -3 2/m"
_symmetry_Int_Tables_number 225
_space_group.reference_setting '225:-F 4 2 3'
_space_group.transform_Pp_abc a,b,c;0,0,0

_atom_site_label
_atom_site_type_symbol
_atom_site_symmetry_multiplicity
_atom_site_Wyckoff_symbol
_atom_site_fract_x
_atom_site_fract_y
_atom_site_fract_z
_atom_site_occupancy
_atom_site_fract_symmform
Mg1 Mg   4 a  0.00000  0.00000  0.00000  1.00000 0,0,0  
Re1 Re   4 b  0.50000  0.50000  0.50000  1.00000 0,0,0  
Ba1 Ba   8 c  0.25000  0.25000  0.25000  1.00000 0,0,0  
O1  O   24 e  0.25951  0.00000  0.00000  1.00000 Dx,0,0 

# end of cif
\end{verbatim}

\normalsize

\bibliography{bibliography}